\documentclass[iop]{emulateapj}
\usepackage[usenames, dvipsnames]{color}
\usepackage[toc,page]{appendix}
\usepackage[section]{placeins}
\usepackage{longtable}

\slugcomment{}
\shorttitle{VVV Survey Microlensing}
\shortauthors{Navarro et al. 2020}

\begin{document}
\title{VVV Survey Microlensing: Catalog of Best and Forsaken Events}

\author{Mar\'ia Gabriela Navarro \altaffilmark{1,2,3,*} 
Rodrigo Contreras Ramos \altaffilmark{3,4}
Dante Minniti \altaffilmark{1,3,5} 
%Etienne Bachelet  \altaffilmark{5} 
Joyce Pullen \altaffilmark{3} 
Roberto Capuzzo-Dolcetta \altaffilmark{2} \&
Philip W. Lucas \altaffilmark{6}
}

\affil{$^1$Departamento de Ciencias F\'isicas, Facultad de Ciencias Exactas, Universidad Andres Bello, Av. Fernandez Concha 700, Las Condes, Santiago, Chile}
\affil{$^2$Dipartimento di Fisica, Universit\`a degli Studi di Roma ``La Sapienza'', P.le Aldo Moro, 2, I00185 Rome, Italy}
\affil{$^3$Millennium Institute of Astrophysics, Av. Vicuna Mackenna 4860, 782-0436, Santiago, Chile}
\affil{$^4$Instituto de Astrof\'isica, Pontificia Universidad Cat\'olica de Chile, Av. Vicu\~na Mackenna 4860, 782-0436 Macul, Santiago, Chile}
\affil{$^5$Vatican Observatory, V00120 Vatican City State, Italy}
\affil{$^6$Centre for Astrophysics Research, University of Hertfordshire, College Lane, Hatfield AL10 9AB, UK}
%\affil{$^5$Las Cumbres Observatory, 6740 Cortona Drive, Suite 102, Goleta, CA 93117 USA }

\begin{abstract}
We search for microlensing events in the zero-latitude area of the Galactic Bulge using the VVV Survey near-IR data. 
We have discovered a total sample of $N=630$ events within an area covering $20.68 deg^2$ between the years 2010 and 2015.
In this paper we describe the search and present the data for the final sample, including near-IR magnitudes, colors
%$K_s$-band light curves, 
and proper motions, as well as the standard microlensing parameters.
We use the near-IR Color-Magnitude and Color-Color Diagram to select $N_{RC}=290$  events with red-clump sources to analyze the extinction properties of the sample in the central region of the Galactic plane.
The timescale distribution and its dependence in the longitude axis is presented. The mean timescale decreases as we approach the Galactic minor axis ($b=0$ deg).
%We perform the photometric and sampling efficiency corrections in order to estimate the total microlensing optical depth.
Finally, we give examples of special microlensing events, such as binaries, short timescale events, and events with strong parallax effect.
\end{abstract}


%%%%%%%%%%%%%%%%%%%%%%%%%%%%%%%%%%%%%%%%%%%%%%%%%%%%%%%%%%%%

\section{Introduction}
Gravitational microlensing effect occurs when a foreground object (e.g planet, star, neutron star, black hole) pass near the line of sight between the earth and a background star causing a deflection of the light rays and generating two images of the source star  \citep{Einstein16, refsdal64, Paczynski86}. 
Due to the resolution of the telescopes available so far, when the effect occurs within our Galaxy, it is not possible to detect both images but an increase in the brightness of the source. 
This effect is useful for a variety of studies in Astrophysics, such as the structure and dynamics of the Galaxy, the search for exoplanet, as a unique way to detect isolated black holes, etc.
The vast applications of microlensing have been extensively reviewed by \cite{Paczynski86, evans03, wam06, gould08, gaudi10, mao12}, among others.

\footnotetext{Corresponding author: mariagabriela.navarro@uniroma1.it}

%The observational study of 
This effect has been studied by Surveys such as Optical Gravitational Lensing Experiment (OGLE; \citealt{Udalski93}), Massive Astrophysical Compact Halo Objects (MACHO; \citealt{Alcock93}), Microlensing Observations in Astrophysics (MOA; \citealt{Bond01}), Exp\'erience pour la Recherche d'Objets Sombres (EROS; \citealt{Aubourg93}), Disk Unseen Objects (DUO; \citealt{Alard95}), Wise Observatory \citep{Shvartzvald12} and Korea Microlensing Telescope Network (KMTNet; \citealt{Kim10}, \citealt{kim17}).
Those surveys have been scanning diverse areas of the sky with high cadence in order to cover the entire range of events, from short timescale events lasting the order of days and suggesting the presence of free-floating planets, to long timescale events of hundreds of days that are good candidates for black holes.
The most interesting areas are those that contain high density of stars because those are the regions where it is more likely to detect a microlensing event. 
Essentially the most explored are the Large and Small Magellanic clouds (LMC, SMC) and the Galactic Bulge (GB). 
The later has been extensively analyzed in detail by the surveys mentioned above excluding the region of lower latitude where the large interstellar extinction hide this area from optical observations, preventing further analysis.

This situation has improved with the emergence of Near IR surveys such as the UKIRT Infrared Deep Sky Survey (UKIDSS; \citealt{Lawrence07}) and the VISTA Variables in the V\'ia L\'actea Survey (VVV; \citealt{minniti10}), that are able to observe throughout the lowest latitude regions of the Galactic plane, reaching well beyond the Galactic center. 

In this context, the study of the most obscured part of the Galaxy using the VVV Survey near-IR data obtained between 2010 to 2015 was first published in \cite{navarro17} covering the 3 innermost tiles of the Survey ($b332$, $b333$ and $b334$). 
\textbf{Then the area was extended covering the whole plane in the Galactic Bulge, that are 14 tiles from $b327$ to $b340$ \citep{navarro18}.
This work presents a complete catalog of the microlensing events found in the innermost area of the Galaxy covering the region within $-10.00^o \leq l \leq 10.44^o$ and $-0.46^o \leq b \leq 0.65^o$. 
The complete study of the microlensing events in the 196 tiles covering the whole Galactic Bulge (area of $300 deg^2$) is a daunting task proposed for the future.}
 
Section \ref{sec:sec2} describes the observations carried out for the VVV Survey and the method used to search for the microlensing events. 
The general analysis of the parameters found for the final sample and the second quality sample is presented in Section \ref{sec:sec3}. 
The Color-Magnitude Diagrams (CMDs), Color-Color Diagrams (CCDs) and the extinction distribution are discussed in Section \ref{sec:sec4}. 
In Section \ref{sec:sec5} we present the spatial distribution of the sample and its dependence on the microlensing timescale. 
Section \ref{sec:sec6} is devoted to the events showing features that depart from the standard model, such as binary lenses or parallax effects. 
In Section \ref{sec:sec7} both, the photometric and sampling efficiency are discussed, and 
in Section \ref{sec:sec8} the timescale analysis is presented. 
The proper motions are discussed in Section \ref{sec:sec9}. 
The relation between our results and the WFIRST observational campaign is mentioned in Section \ref{sec:sec10}.
The comparison between our results with other surveys is presented in Section \ref{sec:sec11}. 
Finally, the conclusions of this work are given in Section \ref{sec:sec12}.

%%%%%%%%%%%%%%%%%%%%%%%%%%%%%%%%%%%%%%%%%%%%%%%%%%%%%%%%%%%%

\section{Observations and Method}
\label{sec:sec2}

The VVV observations are carried out with the VISTA telescope, located at Cerro Paranal - Chile.
This is a 4-m telescope optimized for near-IR observations equipped with the wide-field VISTA InfraRed Camera \citep[VIRCAM]{emerson10} containing 67 million pixels (16 chips of $2048 \times 2048$ pixels). 
The Field of View (FoV) is $1.6 \text{deg}^2$ which is called a ``tile''. 
The entire set of VVV observations comprises 348 tiles; 196 tiles in the Bulge and 152 in the disk area \citep{minniti10, saito12}. 
The VVV observation schedule includes single-epoch photometry in the $Z$ $Y$ $J$ $H$ $K_s$-bands and a variability campaign in the $K_s$-band \cite{minniti10}.

The aperture photometry catalogs for all regions observed by the VVV are produced by the Cambridge Astronomical Survey Unit (CASU) using the VISTA Data Flow System Pipeline \citep{irwin04}.
However, in the area close to the Galactic center where the crowding and differential reddening effects are severe, the PSF photometry is mandatory. 
The reduction of the data for this work was carried out using the DAOPHOT II/ALLSTAR package \citep{stetson87}, as described in detail in \cite{contreras17}. 
The CASU catalogs were used to calibrate our photometry into the VISTA system by means of a simple magnitude shift using several thousands of stars in common.
\textbf{Typical photometric and positional errors are $\sigma_{K_s} = 0.01$ mag, and $\sigma_{J, H} = 0.03$ mag \citep{saito12, contreras17, alonso18} and  $\sigma = 0.1$ arcsec \citep{Libralato15, Smith17, Kurtev17} respectively. }
In this work, the area analyzed comprises 14 VVV tiles in the Bulge (from $b327$ to $b340$) covering the region within $-10.00^o \leq l \leq 10.44^o$ and $ -0.46^o \leq b \leq 0.65^o$. 
The massive analysis of the multi-epoch magnitudes in the $K_s$-band for $\sim 10^7$ point sources was carried out over the past years. 
The reduced dataset included from 73 ($b340$) to 104 epochs ($b333$) spanning six seasons (2010-2015) of observations.

%The standard microlensing model assumes that the source and the lens are point like objects \citep{refsdal64} and describes the total observed flux ($F(t)$) as the combination of the flux of the observed target star ($F_s$) and the background flux ($F_b$) as follows

\textbf{The standard microlensing model assumes that the source and the lens are point like objects \citep{refsdal64} and describes the total observed flux ($F(t)$) as the product of the observed target star ($F_s$) and the microlensing magnification ($A(t)$) as follows}

\begin{equation}
%F(t)  = F_s A(t) + F_b ,
F(t)  = F_s A(t),
\end{equation}

where $A(t)$ can be expressed as

\begin{equation}
A(t) = A(u(t)) = \frac{u^{2} +2}{u \sqrt{u^{2} +4}},
\end{equation}

and $u$ is the projected separation of the source and the lens in units of Einstein radius and is related to the impact parameter and thus with the amplitude of the light curve. 
The variation of $u$ with time is described by

\begin{equation}
u (t) = \sqrt{u_0^2 + \left( \frac{t-t_0}{t_E} \right)^2}.
\end{equation}

Where $u_0$ is the minimum impact parameter, $t_0$ the time of maximum magnification and $t_E$ the Einstein radius crossing time.\\

\textbf{In crowded fields, the seeing element of the sources can be affected by the light of nearby stars. 
In these cases the observed microlensing magnification ($A(t)^{obs}$) follows}

\begin{equation}
[A(t)^{obs} -1] = [A(t) -1] f_{bl} ,
\end{equation}

\textbf{where $f_{bl}$ is the blending parameter and $f_{bl}=1$ for the cases of no blending and $f_{bl} \sim 0$ for extreme blending.}

The microlensing event selection follows the same procedure outlined by \cite{navarro17}. 
First of all, from the complete sample ($\sim 10^7$ light curves) we selected the light curves with more than 20 data points, so the sample was reduced to $\sim 5 \times 10^6$ . 
The second step was to perform an automated fitting procedure using the standard microlensing model.
Using relations between the data points based in the reconstruction of a microlensing light curve, i.e. the detection of the appreciable increase followed by a decrease in brightness, we got an initial guess for the standard microlensing model parameters $u_{0}$, $t_{0}$ and $t_E$.
The approximate time of maximum magnification ($t_0$) is defined as the moment when the brightness starts to decrease 
and the timescale ($t_E$) to the amount of time from when the brightness increase begins until it returns to its normal state (divided by two).

After the fitting procedure, the quality index ($q$) that tells us the probability that the light curve is a microlensing event is computed.
The value of this index varies between 0 and 1 so that the light curves that are most likely to be a microlensing event have low indices.
The features that decrease the quality number (i.e increase the probability that the event is a microlensing event) are; 
the number of data points during the event with a minimum number of 4 data points;
the number of data points in the baseline in order to well constrain the baseline magnitude and avoid confusion with long period variables; 
the dispersion of the residuals between the fitted light curve and the data during the event to evaluate if the event follow the point source point less (PSPL) model (which is the most important indicator) and
the dispersion of the residuals between the fitted light curve and the data in the baseline.
Each indicator is included in the calculation of the quality index with different weight according to its relevance.
\textbf{The quality number was computed as the multiplication of 1 over every feature as follows: }
\begin{equation}
q = \text{weight}_i (1/\text{feature}_i) \times {\text{weight}_{i+1}} (1/\text{feature}_{i+1})...
\end{equation}
After the fitting procedure and the estimation of the quality index we evaluated four criteria: 

\begin {enumerate}
\item Binaries
\item Timescale
\item Magnitude amplitude
\item Quality index
\end {enumerate}

In order to search for events following the PSPL model we limited our sample to the light curves with constant increase and decrease in the magnification within a specific period ($t_E$ initial guess) and non-decreases in brightness within this time in order to exclude long timescale binary events and strong parallax events. 
We rejected the light curves with initial guess in timescale larger than 500 days, amplitude $<0.1$ and quality index $q >0.002$. 
With amplitude we refer to the $\Delta$ mag between the baseline magnitude and the brightest data point.
Applying this criteria we decrease the initial sample to $\sim 5 \times 10^4$ light curves (from $2500$ to $5000$ per tile).
%The fitting procedure was also performed using the unblended approach in order to sanity check our results. 
%However, due to the crowding that is critical in this region, all analysis was done using the model including blending.

The third step was a double blind visual inspection of all the microlensing light curves and models fitted, to apply a series of customized requirements (same requirements applied in \cite{navarro17}), such as: 
a well-defined constant baseline; 
at least one data point in the rising and falling light curve; 
at least four points with $4\sigma$ above the baseline; 
baseline covering more than one season; 
timescales within an acceptable range to avoid confusion with long period variable stars, 
and a good fit to the microlensing standard model.
The strict criteria applied here ensure that the parameters obtained are well constrained.

After this selection procedure the final sample was reduced to 655 events. 
In this step we also selected a second quality sample with forsaken events that do not satisfy at least one of the criteria even though many of them show a clear microlensing event light curve. 
Additionally, we flagged some events that show clear binary source (binary lens) light curves, and parallax effects, that depart from an appropriate fit to the microlensing curve. 
The number of light curves analyzed in each step along with a short description are shown in Table~\ref{tab:table0}.

The observational strategy of the VVV Survey implies 6 individual observations (``pawprints'') with small offsets in order to fill the gaps between the 16 detectors of VIRCAM \citep{catelan11}. 
Considering that the detection procedure was performed for the pawprints and not for the complete tile, there is a small probability to obtain duplicated detections. 
These are very useful as an external test of the quality of the various parameters measured from our experiment. 
In order to select these duplicate events, we searched for pairs of events that meet the following 3 criteria simultaneously; 
distance differences less than 2 arcsec; 
$t_0$ differences less than 7 days 
and baseline $K_s$ magnitude differences less than $0.15$ mag.
In this case we included the forsaken sample because the duplicated events will have twice the data points and can eventually fulfill all the requirements mentioned above. 
We found 30 duplicated events, that is the $\sim 5\%$ of the complete sample.

The final sample contains 630 microlensing events (\cite{navarro17}, \cite{navarro18}). 
\textbf{As an additional check we analyze the 630 light curves during the efficiency analysis discussed in Section \ref{sec:sec7}. 
The final sample was recovered as good candidates as microlensing events. }

%%%%%%%%%%%%%%%%%%%%%%%%%%%%%%%%%%%%%%%%%%%%%%%%%%%%%%%%%%%%

\section{Search for Microlensing Events}
\label{sec:sec3}
As mentioned in Section~\ref{sec:sec2} the final sample is divided in two sets, the 630 first quality events which fulfill all the conditions, and 2010 lower quality forsaken events.

\subsection{Catalog of Best Microlensing Events}
The first quality catalog of the 630 microlensing events is listed in Table~\ref{tab:table1}, including their positions in Galactic coordinates, $K_s$-band magnitudes, and the parameters obtained by the standard microlensing model fitting procedure along with the respective errors.

\begin{figure}[t]
\epsscale{1.2}
\plotone{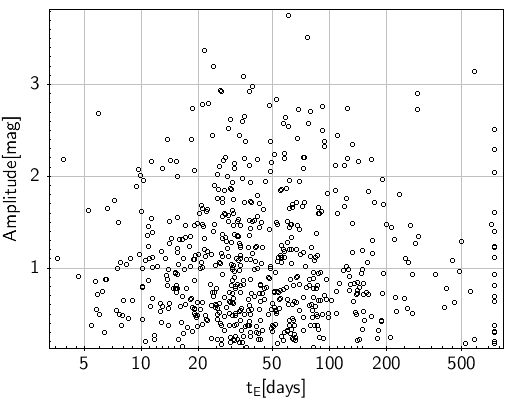}
\caption{Amplitude vs Einstein radius crossing time $t_E$ for the 630 first quality events. Both parameters were obtained using the standard microlensing model. \\
\label{amp_timescale}}
\end{figure}

In order to check the fidelity of our results we analyzed the parameters obtained by the standard microlensing model. 
Figure~\ref{amp_timescale} shows the dependence of the microlensing amplitude with the timescale. 
The distribution does not show a significant trend but a homogeneous distribution, as expected. 

Some high magnificationation events are observed ($N=6$ events with Amp$>3$ mag), but the low magnification events clearly dominate the sample. 
There are many long timescale events ($t_E>100$ days), including $N=49$ events with $t_E>200$ days
There are also short timescale events ($t_E<10$ days), including $N=3$ events with $t_E<5$ days.
The events with long timescale are specially interesting because they have the potential of detecting Primordial Intermediate-Mass Black Holes, with masses between $20$ and $10^5 M_\odot$ \citep{frampton16}.

\begin{figure}
\epsscale{1.2}
\plotone{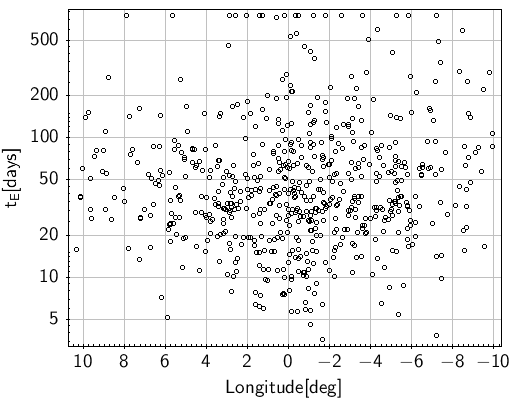}
\caption{Distribution of the Einstein radius crossing time $t_E$ vs the Galactic longitude for the first quality sample. \\ 
\label{longte}}
\end{figure}

As an additional check we analyzed the timescale (in logarithmic scale) versus longitude relation (see Figure~\ref{longte}). 
This diagram shows a homogeneous distribution without significant variations in the mean timescales with Galactic longitude, as expected. 

\subsection{Catalog of Forsaken Microlensing Events}
\begin{figure}[t]
\epsscale{1.2}
\plotone{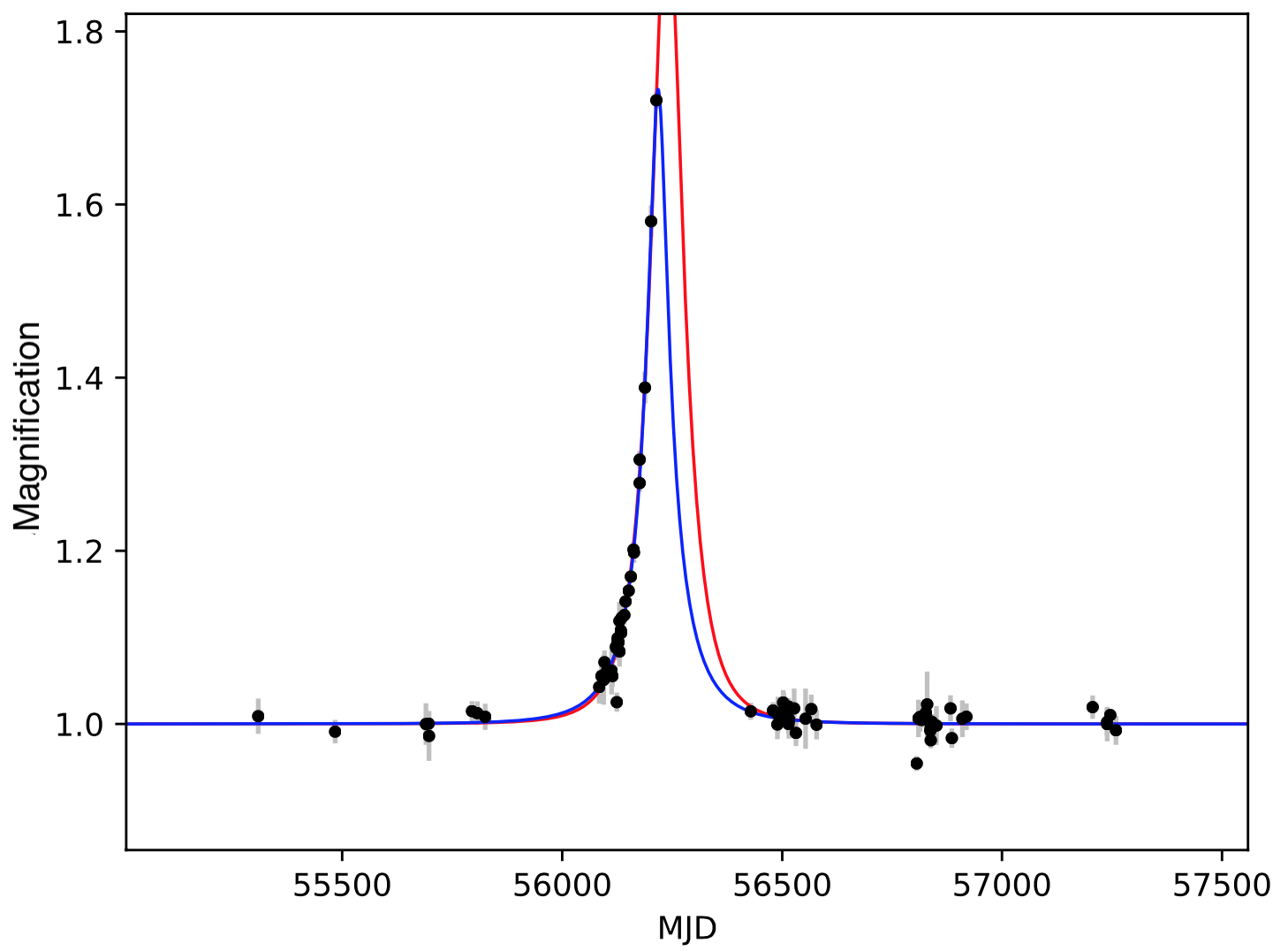}
\caption{Examples of a forsaken event light curves. The lack of data points in the decreasing part of the curve does not allow us to calculate the Einstein radius crossing time $t_E$ properly. \\
\label{LCforsaken}}
\end{figure}

The events that fail to fulfill  at least one of the criteria defined in the previous section are also selected and inspected.
Visual inspection of the light curves reveals that some of them are likely to be microlensing events and therefore we single them out as second best (or forsaken) microlensing events. 
A representative light-curve is shown in Figure~\ref{LCforsaken}. 
It is clear that the event follows the standard microlensing model but do not meet all the requirements. 
In the most evident cases, the lack of one data point on the rising or falling area of the light curve prevents us from obtaining an accurate estimate for the Einstein radius crossing time.
Indeed, the efficiency analysis that follows predicts that we are missing many real events and some of them may be these forsaken events listed here. 
Table~\ref{tab:table2} presents the catalogue of the 2010 forsaken microlensing events. 
However, they are listed here only for completeness (and because it would be a shame that they are entirely lost), but they are not considered any further for the analysis of the events efficiency, distribution and timescale.

%As with the first quality sample, we analyzed the relation between the amplitude and timescale and the longitude dependence of the timescale in a logarithmic scale (Figures~\ref{amptimefor}, ~\ref{longtfor}). 
%Both figures show that there are no trends as expected.

\begin{figure}
\epsscale{1.2}
\plotone{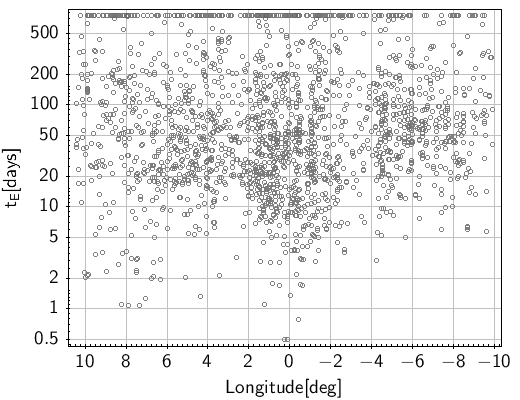}
\caption{Distribution of the Einstein radius crossing time $t_E$ vs the longitude for the 2010 forsaken events. \\ 
\label{longtfor}}
\end{figure}

We analyzed the relation between the longitude dependence of the timescale in a logarithmic scale (Figure~\ref{longtfor}). 
From Figure~\ref{longtfor} we see that there are no trends as expected.

%\begin{figure}[t]
%\epsscale{1.2}
%\plotone{amp_timescalefor.png}
%\caption{Amplitude vs timescale for the 2010 forsaken events. \\
%\label{amptimefor}}
%\end{figure}

%%%%%%%%%%%%%%%%%%%%%%%%%%%%%%%%%%%%%%%%%%%%%%%%%%%%%%%%%%%%

\section{Color-Magnitude Diagrams}
\label{sec:sec4}
For purposes the timescale analysis, it is profitable to use a subsample of Red Clump Giants stars (RC) \citep{Paczynski98}. 
Red Clump Giants stars are evolved low mass stars that are burning helium in their cores and are located in a narrow band (Horizontal Branch) of the CMD, hence act as standard candles and excellent distance indicators. 

There are two main reasons to analyze specifically the RC sample; 
the probability of the source is located in the Bulge is high and they are bright enough that the blending effect might be negligible.
Therefore the parameters obtained from the light curve fits are more reliable \citep{pop01}. 

\begin{figure}[t]
\epsscale{1.2}
\plotone{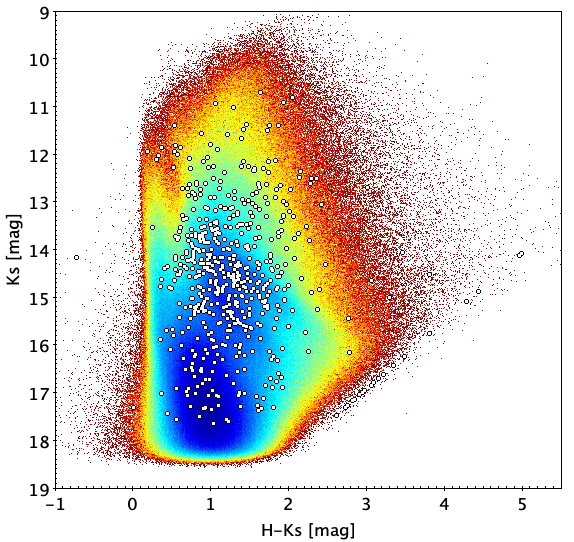}
\caption{Near-IR $K_s$ $vs$ $H-K_s$ Color-Magnitude Diagram for the 14 VVV tiles (from $b327$ to $b340$). 
The white circles indicate the sources of the first quality sample of microlensing events. 
The events concentrated at the red edge of the distribution have $J$-band values below the detection limit. So the color was computed as a lower limit using the CMD.\\
\label{cmdh}}
\end{figure}

Figure~\ref{cmdh} shows the $K_s$ vs $H-K_s$ CMD of the area studied here, plotted as a density map along with the sources of the microlensing events.  
The CMD of the $K_s$ vs $J-K_s$ is shown in Figure 1. of \cite{navarro18}. 
The RC sources were selected using the $K_s$ vs $J-K_s$ CMD. 
Its clear that the RC stars are extremely affected by the differential extinction present in the inner Bulge and plane and also that they follow the extinction law of \cite{alonso17} computed for the VVV data in the most central fields with a slope of $A_k / E(J-K_s) = 0.428 \pm 0.005 \pm 0.04$ and $A_k / E(H-K_s) = 1.104 \pm 0.022 \pm 0.2$ for each CMD.
The CMD of the forsaken events (not shown for the sake of space) is similar to that of the CMD of bonafide events shown in Figure~\ref{cmdh}.

\begin{figure}
\epsscale{1.2}
\plotone{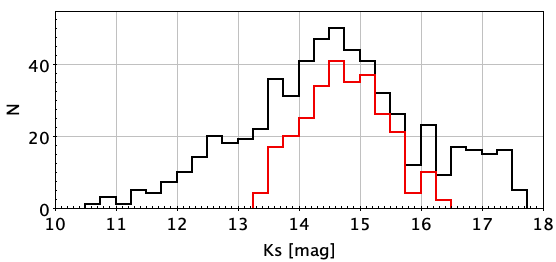}
\caption{$K_s$-magnitude distribution for the detected events (black histogram) compared with the selected RC events (red histogram).
The events concentrated at the red edge of the distribution have $J$-band values below the detection limit. So the color was computed as a lower limit using the CMD.\\
\label{KHist_log}}
\end{figure}

The $K_s$ magnitude histogram is presented in Figure~\ref{KHist_log} in order to show the red clump selection. 
The over density of sources centered in $K_s= 14.5$ mag reveals the presence of the red clump sources. 
The peak of the distribution matches the reddened magnitudes of the RC giants located at different distances. 
% \textcolor{red}{Menciono algo del Bump in kmag 17? \\}

\subsection{Wesenheit Color-Magnitude Diagrams}
In order to identify better the RC in de CMD of areas where the extinction is extremely high but also variable such as in the Galactic Bulge,
it is more advantageous to use the reddening corrected Wesenheit magnitude that is defined as: 

\begin{equation}
W_{k_s} = K_s - 0.428 (J-K_s),
\end{equation}

\begin{figure}[t]
\epsscale{1.2}
\plotone{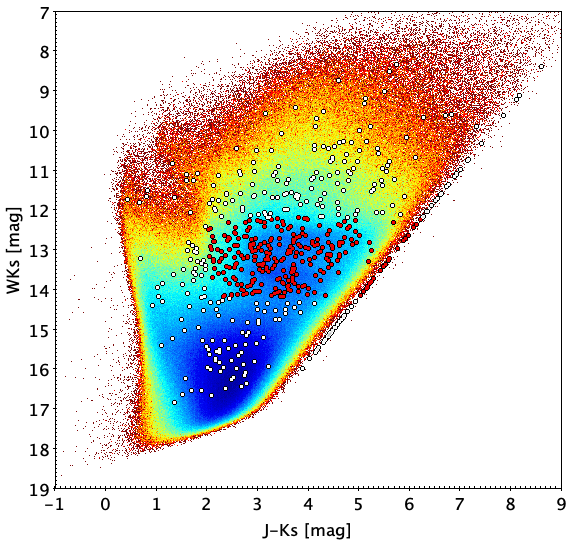}
\caption{Reddening independent Wesenheit $W_{k_s}$ magnitude vs $J-K_s$ diagram for the 14 VVV tiles (from $b327$ to $b340$). 
The white circles are the first quality microlensing events found and the red circles indicate the sources of the sample located in the Red Clump.\\
\label{cmdw}}
\end{figure}

The $W_{k_s}$ vs $J-K_s$ CMD is shown in Figure~\ref{cmdw}. 
The distribution of sources shows the deep near-IR VVV Survey detection limit of $W_{k_s} \approx 17.5$ mag ($\sim K_s \approx 17.5$ mag). 
This is much deeper than the other surveys dedicated to search for microlensing events in this region, suggesting that our sample can include events with sources located in the far disk. 
Assuming that the RC sources are located within the following limits: $12.0<W_{k_s}<14.0$ mag, and $(J-K_s) >2$ mag, we obtained 290 RC sources in total. 
This amounts to a large fraction ($46\%$) of our total sample of bonafide microlensing events.

In the CMDs shown in Figure~\ref{cmdh} and Figure~\ref{cmdw}, the events located at the red edge of the distribution have a $J$-band value below the detection limit. 
Thus, the limiting $J$-band magnitude for each $K_s$ and $W_{k_s}$ is assumed to obtain the color for these cases. 

\subsection{Color-Color Diagrams}
\begin{figure}[t]
\epsscale{1.2}
\plotone{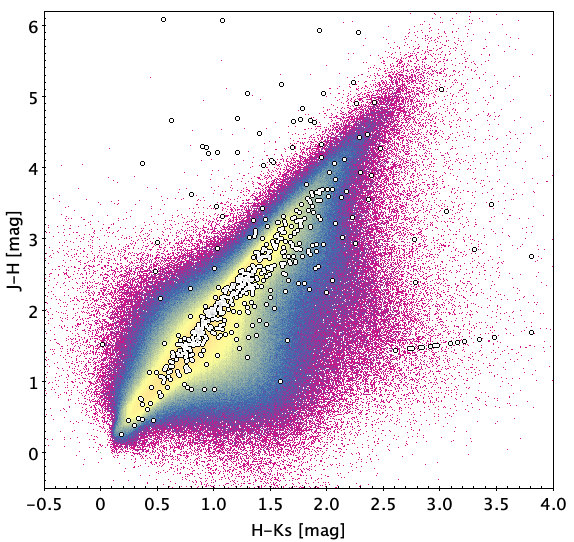}
\caption{Color-Color diagram of the 14 VVV tiles as a density map along with the 630 microlensing events found (white circles).
The line of events at colors $(H-K_s)>2$ mag have $J$-band values below the detection limit. So the $J-K_s$ color was computed as a lower limit using the  $K_s$ $vs$ $J-K_s$ CMD. \\
 \label{colorcolor}}
\end{figure}

The Color-Color diagram (CCD) for the first quality microlensing events is presented in Figure~\ref{colorcolor}.
%where the color-coding is the same as in Figure~\ref{cmdh} and Figure~\ref{cmdw}. 
This Figure shows that most sources have large reddening, in agreement with them being located in the Bulge and beyond. 
The trend at colors $(H-K_s) >2$ mag is an artefact due to the detection limit, and to the assumption used to obtain a representative value for the sources in $J$ and $K_s$. 

\subsection{The Extinction Distribution}
\begin{figure}[t]
\epsscale{1.2}
\plotone{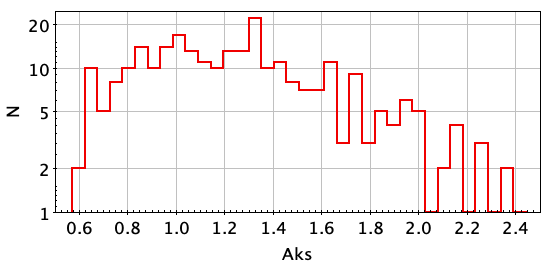}
\caption{Extinction distribution for the sources located in the Red Clump.
The distribution is within the values of the reddening map for the same area from \cite{gonzalez12}.
 \\
 \label{ext}}
\end{figure}

For the sample of RC giants, we can estimate individual extinction values assuming  the mean intrinsic magnitude and color of the RC to be $K_{s0}= -1.68\pm 0.03$ mag, and $(J-K_s)_0=0.60\pm 0.01$ mag \citep{alves02}. 
We adopt the extinction ratios for this region from \cite{alonso17}: $A_{K_s}=0.428 \times E(J-K_s)$. 
Figure~\ref{ext} shows the extinction distribution for the Red Clump sources of the first quality sample.
The distribution is within the values of the reddening map for the same area from \cite{gonzalez12}.

We make a blue color cut in order to select distant RC giants. 
The limits at $(J-K_s)>2.0$ mag ($(H-K_s)>0.3$ mag), yields extinction and reddening values of $E(J-K_s)>1.40$ and $A_{K_s}>0.60$ mag, respectively.
This ensures that we are dealing with distant RC stars, located beyond the great Galactic dark lane.
That is an optically thick dust lane located at low latitudes, covering the Galactic plane fields studied here across the Bulge with $-10^o<l<10^o$  \citep{minniti16}.
This great dark lane causes a mean extinction of $A_V> 1.8$ mag, and all the RC events are located beyond this feature because the blue color cut yields $A_V>6$ mag approximately. 

The reddest RC sources detected in both the $J$ and $K_s$-bands have $(J-{K_s})=5.0$ mag implying $E(J-{K_s})=4.40$, and $A_{K_s}=1.88$ mag.
However, some of the most reddened sources are not detected in the $J$-band, and we have to use the $H$-band. 
The reddest RC sources detected in both the $H$ and $K_s$-bands have $(H-K_s)\sim 2.3$ mag yielding $A_{K_s}=2.36$ mag.
Thus, the reddest sources of microlensing events observed here have optical extinctions up to $A_V\sim 21$ mag. 
Such microlensing events are beyond detection for current optical surveys, and only a microlensing search with the Wide Field Infrared Survey Telescope (WFIRST; \citealt{Green12}, \citealt{Spergel15}) in this region would be capable of improving upon the present results.

%%%%%%%%%%%%%%%%%%%%%%%%%%%%%%%%%%%%%%%%%%%%%%%%%%%%%%%%%%%%

\section{Spatial Distribution}
\label{sec:sec5}
\begin{figure*}[t]
  \includegraphics[width=\textwidth]{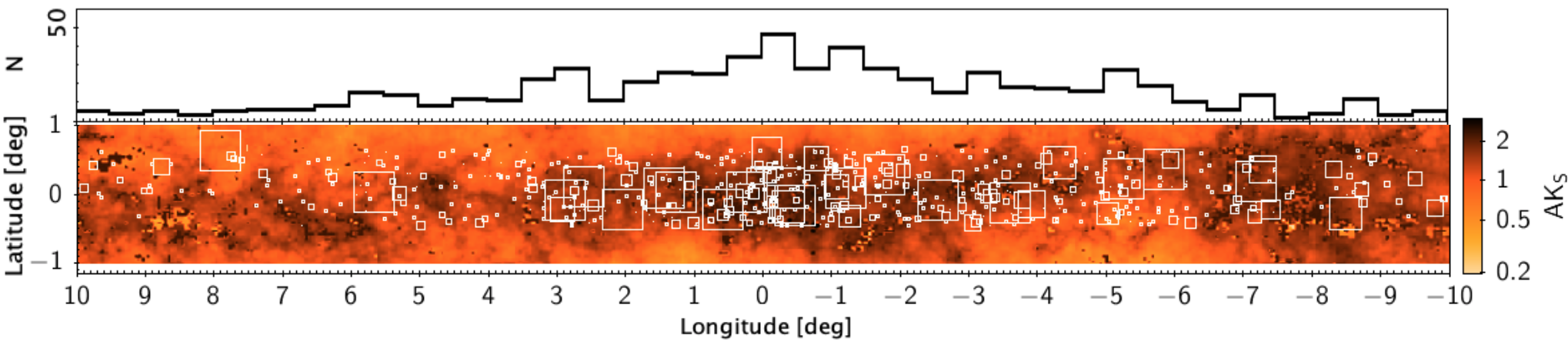}
\caption{Map showing the location of the VVV microlensing events (white squares) around the Galactic center, overlaid on the extinction map of \cite{gonzalez12}. 
The square sizes are proportional to the event timescales in logarithmic scale. \textbf{The histogram showing the number of events and its dependence with longitude is shown on top.}
The duplicate events in the overlapping areas have been accounted for. \\
 \label{Spatial_Distribution_W}}
\end{figure*}

\begin{figure*}[t]
 \includegraphics[width=\textwidth]{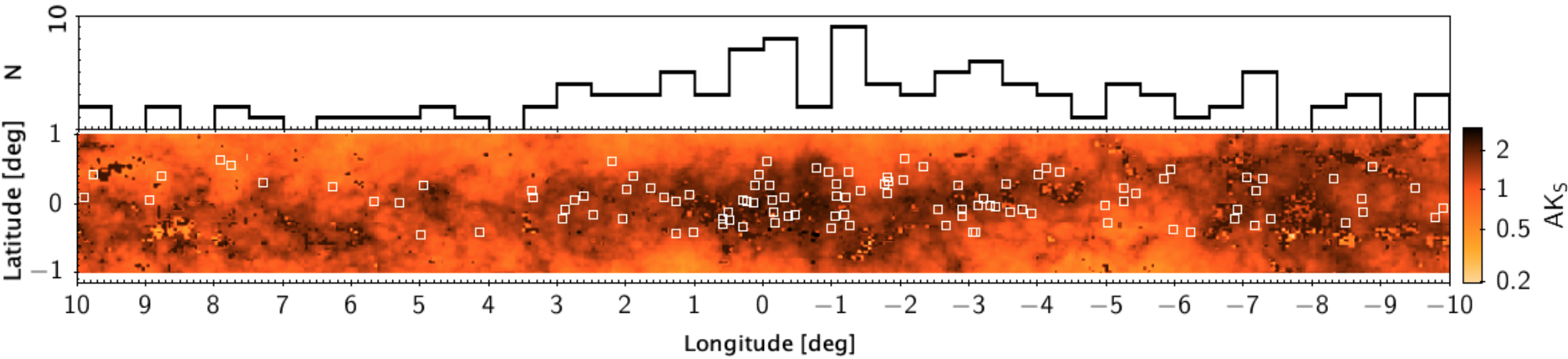}
\caption{Spatial distribution map for the long timescale VVV microlensing events ($t_E>100$ days), overlaid on the extinction map of \cite{gonzalez12}.  \textbf{The histogram showing the number of events and its dependence with longitude is shown on top.}
\label{spatiallong2}}
\end{figure*}

\begin{figure*}[t]
  \includegraphics[width=\textwidth]{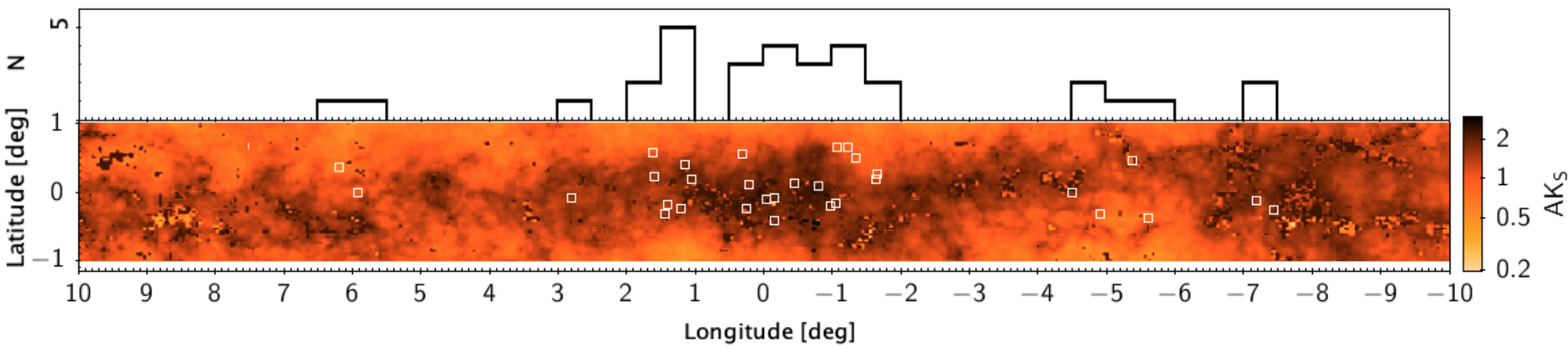}
\caption{Spatial distribution map for the short timescale VVV microlensing events ($t_E<10$ days), overlaid on the extinction map of \cite{gonzalez12}.  \textbf{The histogram showing the number of events and its dependence with longitude is shown on top.}\\
\label{spatialshort2}}
\end{figure*}

The global spatial distribution of the best event sample is presented in \cite{navarro18}. 
In order to complement that work, we show in Figure~\ref{Spatial_Distribution_W} the distribution of the events overlaid in the extinction map of \cite{gonzalez12}. 
The square sizes in Figure~\ref{Spatial_Distribution_W},~\ref{spatiallong2} and ~\ref{spatialshort2} are proportional to the Einstein radius crossing times in logarithmic scale. 
It is possible to identify a wide range of timescales.
The distribution is uniform but a slight concentration towards negative latitude is detected \citep{navarro18}. 
%A clear concentration of long timescale events near the center of the Galaxy is detected.
Also, we find an excess of events in the new window of low extinction discovered by \cite{oscar18}.

The spatial distribution maps for the long (with $t_E>100$ days)  and short timescale events (with $t_E<10$ days) are shown in Figure~\ref{spatiallong2} and Figure~\ref{spatialshort2}, respectively. 
From the histogram in the top of each figure we see that the long timescale events appear to be more uniformly distributed with longitude, while on the other end, the shortest timescale events appear more concentrated to the Galactic center. 
Additionally, the long timescale events exhibit an asymmetric distribution, with twice the number of events at negative longitudes. 
Indeed, this effect was predicted by the models of \cite{han95}, as a consequence of the barred Bulge.
However, these apparent trends are based on small numbers of events and need to be confirmed with larger samples.

\begin{figure}
\epsscale{1.2}
\plotone{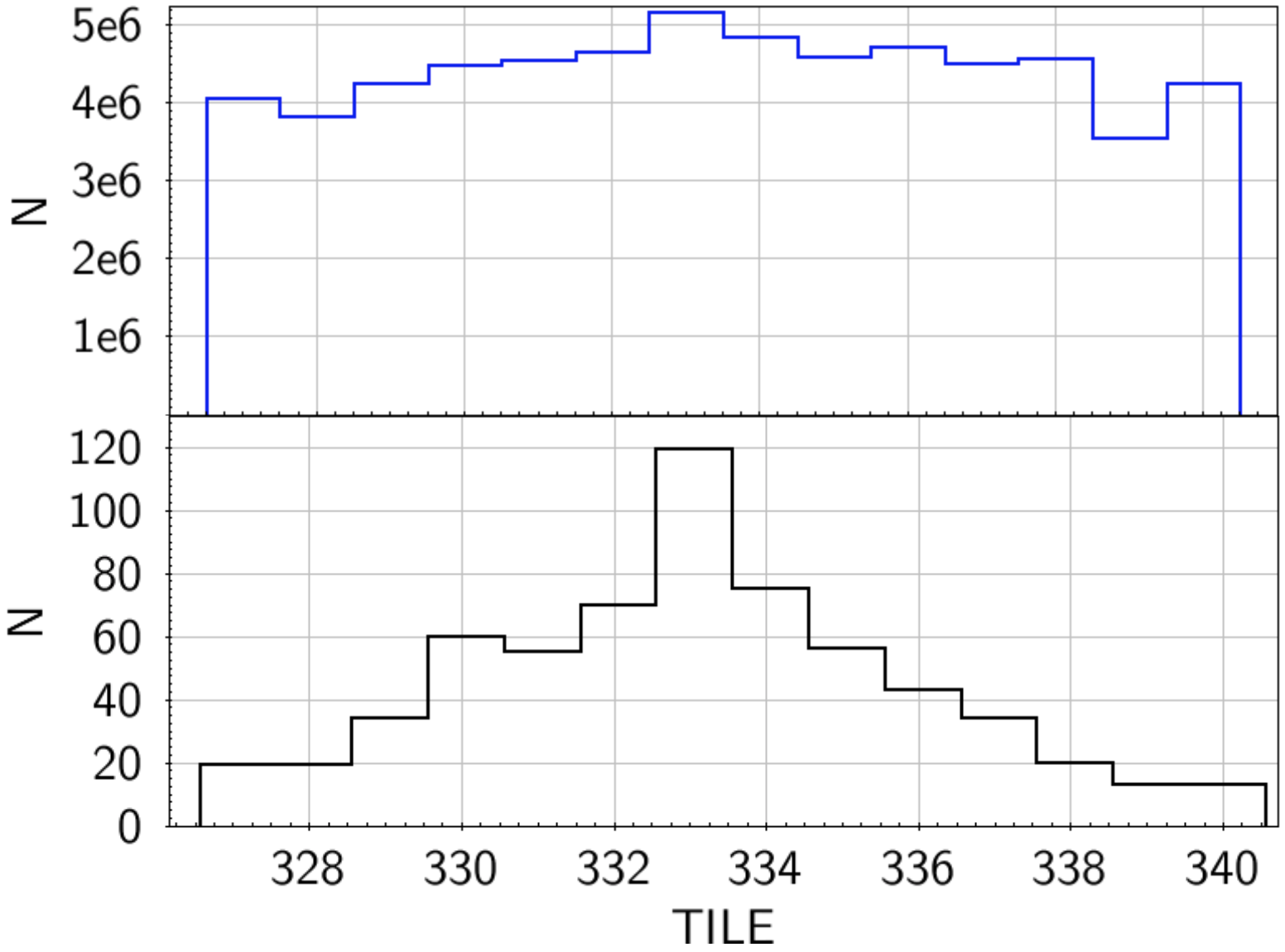}
\caption{Upper panel: Number of light curves analyzed per tile. 
Lower panel: Number of events detected per tile. 
Note that the tile IDs increase with increasing Galactic longitude (i.e. positive longitudes are located to the right).\\ 
\label{nstars}}
\end{figure}

The number of sources analyzed and events detected in each tile are presented in Figure~\ref{nstars}. 
The number of sources analyzed (upper panel) slightly increase toward the central tiles ($b332$, $b333$ and $b334$). 
The same behavior is seen for the events detected (lower panel), but the increase is more prominent. 
This means that the increase of microlensing events is not an effect just due to the increase in the density of stars but related to an increase of lenses in this direction.

%%%%%%%%%%%%%%%%%%%%%%%%%%%%%%%%%%%%%%%%%%%%%%%%%%%%%%%%%%%%

\section{Special microlensing cases}
\label{sec:sec6}
There are several special microlensing events that depart from the usual standard microlensing light curve model. 
The detection of these special cases underscores the ability of any microlensing survey to recover good events.

\subsection{Binary events}
Binary events are special cases of microlensing that involve binary lenses or sources (e.g. \cite{mao91}, \cite{griest93},  \cite{Alcock20}, \cite{Skowron09}).
A binary lens event occurs when the lens is in a binary (or a multiple system).
Binary lens events exhibit a large variety of light curves, and these include the special cases of extrasolar planets.
A binary source event occurs when the source is in a binary system. 
In this case, the light curve is a simple superposition of the light curves of the two sources.

The microlensing surveys in general are not designed to find binary events because of their wide variety of light curve shapes. 
Anyway, there are new tools to deal with such events like PyLima \citep{bachelet17}. %specially designed for the detection of extrasolar planets with the WFIRST.
Binary events contain additional information (relative proper motions) that potentially allows to break the degeneracy between masses and distances.
However, they require additional parameters to fit the light curves, and in some cases, there is ambiguity where more than one fit can be acceptable.

\begin{figure}
\epsscale{1.2}
\plotone{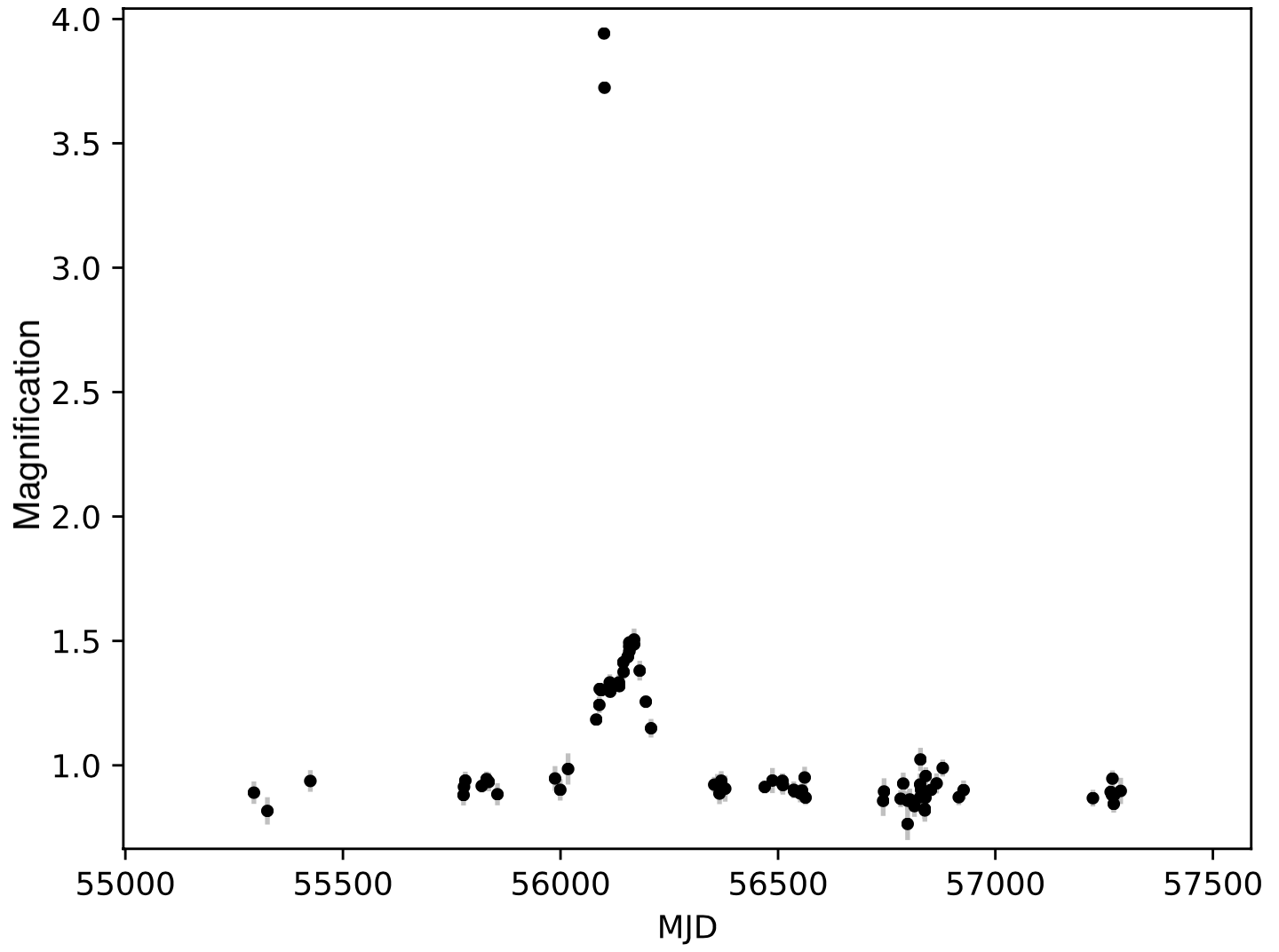}
\caption{Example of a VVV binary microlensing event. 
The system is not a single star but a binary with one (main peak) more massive than the other (second shorter peak). \\
\label{binaryevent}}
\end{figure}

During the course of our survey we have discovered several binary events.
Figure~\ref{binaryevent} presents an example of a VVV binary microlensing event. 
The short timescale curve overlaid in a low amplitude long microlensing event light curve suggest the presence of a low mass companion.
Their detailed study is beyond the purposes of the present work, but they also illustrate the VVV survey ability to detect binary microlensing.

\subsection{Short timescale events}
Short timescale events are very interesting special cases of microlensing because they can potentially single out very low mass objects like late-type stars, brown dwarfs, and even free-floating planets.

\begin{figure}
\epsscale{1.2}
\plotone{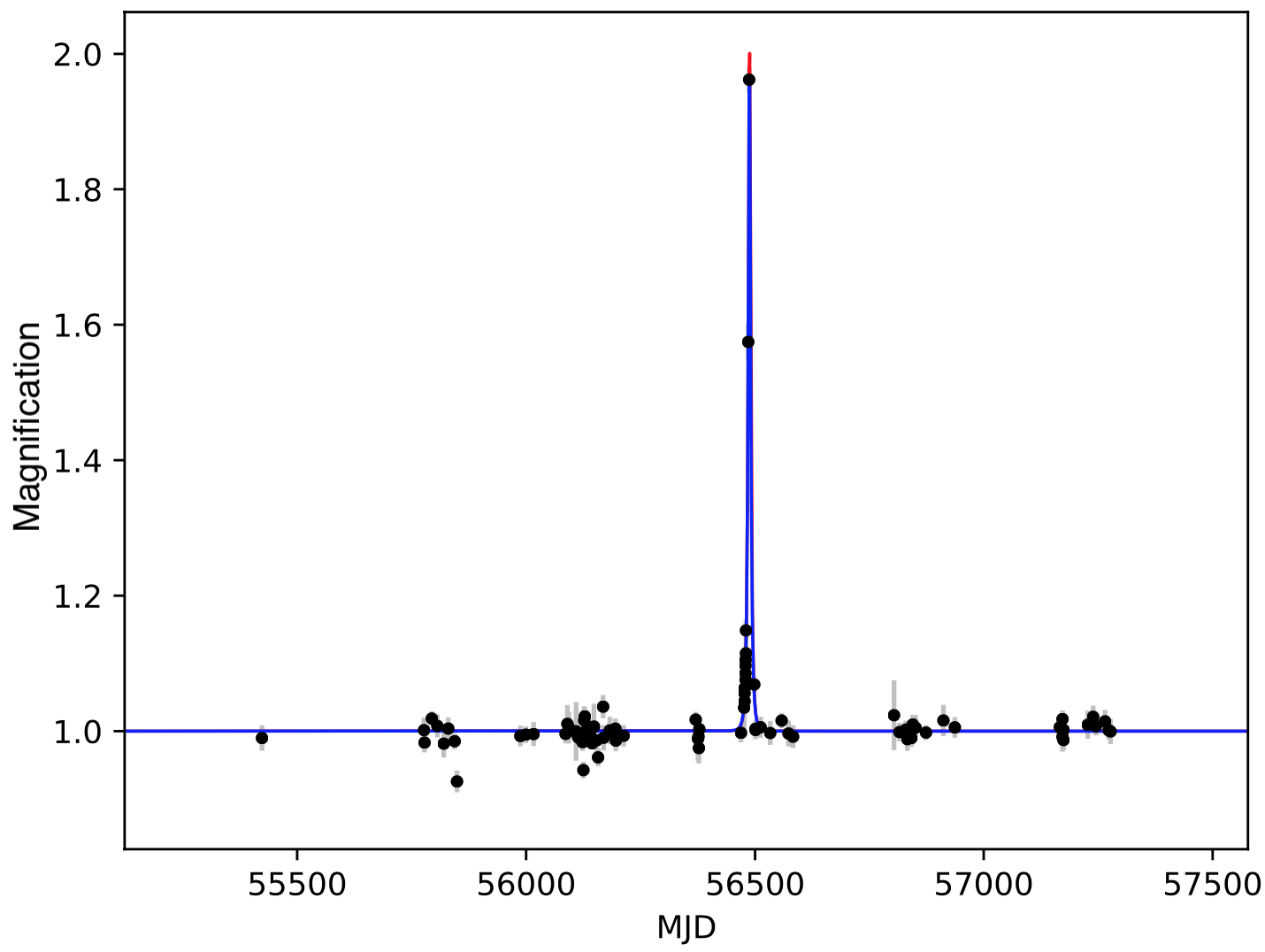}
\caption{Example of a short timescale microlensing event with the standard model fitting. 
The event has $>10$ observations during the event making it possible to constrain the parameters. \\
\label{shortevent}}
\end{figure}

The VVV survey should be almost insensitive to very short timescale microlensing events because the cadence of the observations is not very high (nightly at best). 
However, we have still detected a number of them. 
The detailed study of this population is not optimal with the VVV sample due to the low efficiency that VVV has specially for short timescale events.
Figure~\ref{shortevent} presents a case of a VVV short timescale event.
Such events suggest the presence of free-floating planets but due to the degeneracy with the distance and the transverse velocity, further analysis is needed to confirm this.

\subsection{Parallax events}
\begin{figure}
\epsscale{1.2}
\plotone{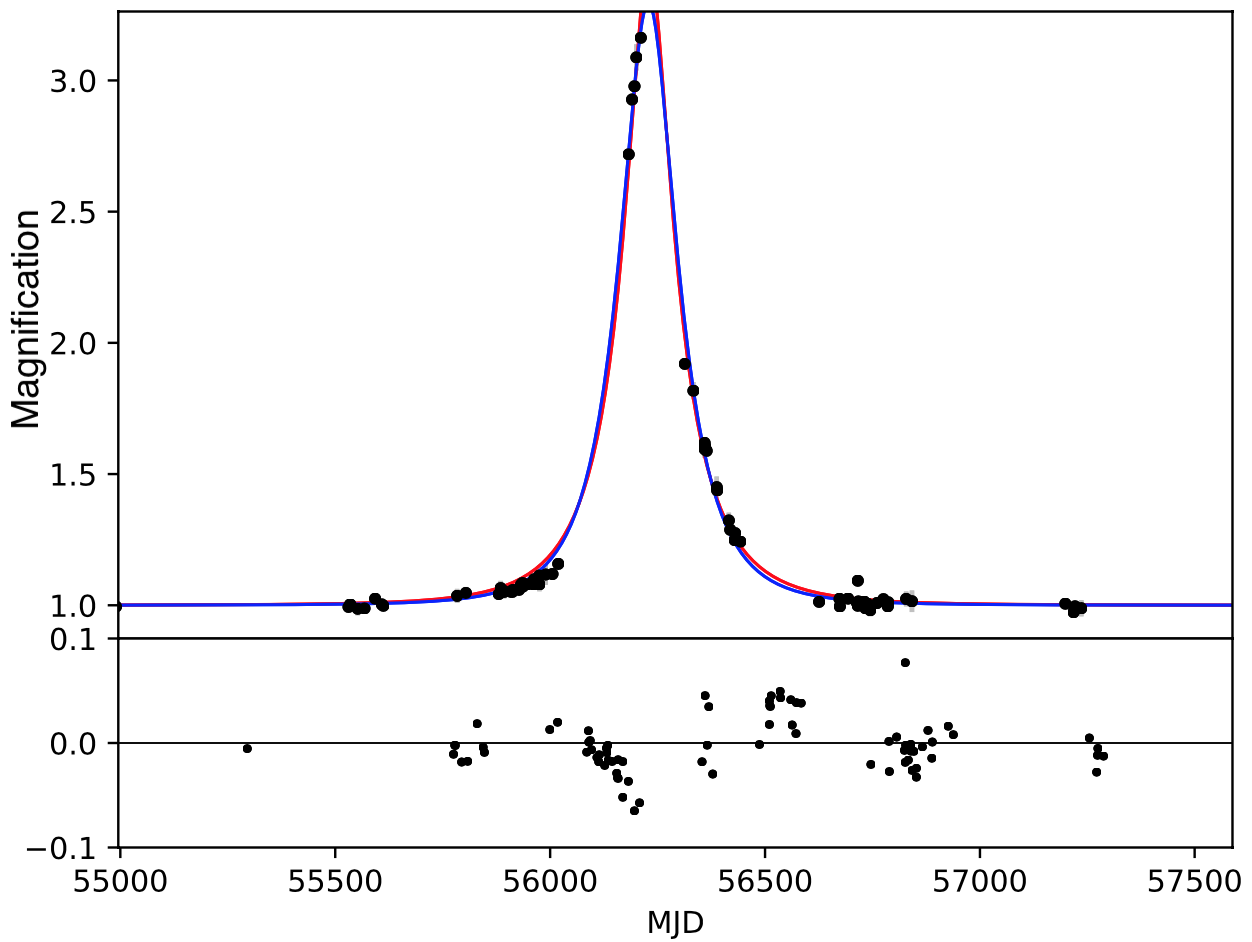}
\caption{Upper panel: Example of a parallax event light curve along with the PSPL model fit.
Lower panel: The residuals of the PSPL fit, showing the expected behavior for this kind of events. \\ 
\label{parallax}}
\end{figure}

Microlensing events with parallax effect are very important special cases of microlensing because they can allow to break the microlensing degeneracy,  solving for the mass of the lens (e.g. \cite{Alcock95}, \cite{Alcock20}, \cite{gould92}, \cite{Bennett1998}).
This effect is detectable generally in long timescale events, where the shape of the microlensing light curve is slightly altered due to the motion of the Earth on its orbit around the Sun.
An alternative case occurs when the source is a binary system and therefore exhibits an orbital motion (this is called the reverse parallax or xallarap effect).
Figure~\ref{parallax} presents an example of a VVV microlensing event with evident parallax effect. 
The standard microlensing model is not able to fit the event due to the asymmetry that is characteristic of this events. 
The residuals of the fit to the standard microlensing light curve show the typical effect due to the parallactic motion of the Earth.

%%%%%%%%%%%%%%%%%%%%%%%%%%%%%%%%%%%%%%%%%%%%%%%%%%%%%%%%%%%%

\section{Total Microlensing Efficiency}
\label{sec:sec7}
The efficiency of the survey detecting microlensing events should be considered for the correct analysis of the timescale distribution. 
Following \cite{Alcock21}, we explore the total microlensing efficiency. 
We study the first approximation efficiency of our experiment by dividing into photometric efficiency and sampling efficiency and treating them independently.

\subsection{Photometric Efficiency}
The photometric efficiency is taken from artificial star simulations for the RC located in the VVV tiles mapped here. 
In this area the completeness ranges between $40\%$ to $75\%$, with the tiles located at negative Galactic longitudes being more incomplete (See Fig. 1 of \cite{elena16} and \cite{surot}). 
The PSF photometry used in this study is explained in greater detail by \cite{alonso18}. 
We acknowledge the limit of our analysis, the photometric efficiency correction can be applied to the sources brighter than the RC, i.e., $K_s \leq14$ mag and in this case this correction increase the gradient found here for both the complete sample and the RC sources, predicting that there are many more microlensing events at negative than at positive Galactic longitudes. 

\subsection{Sampling Efficiency}
\begin{figure}
\epsscale{1.2}
\plotone{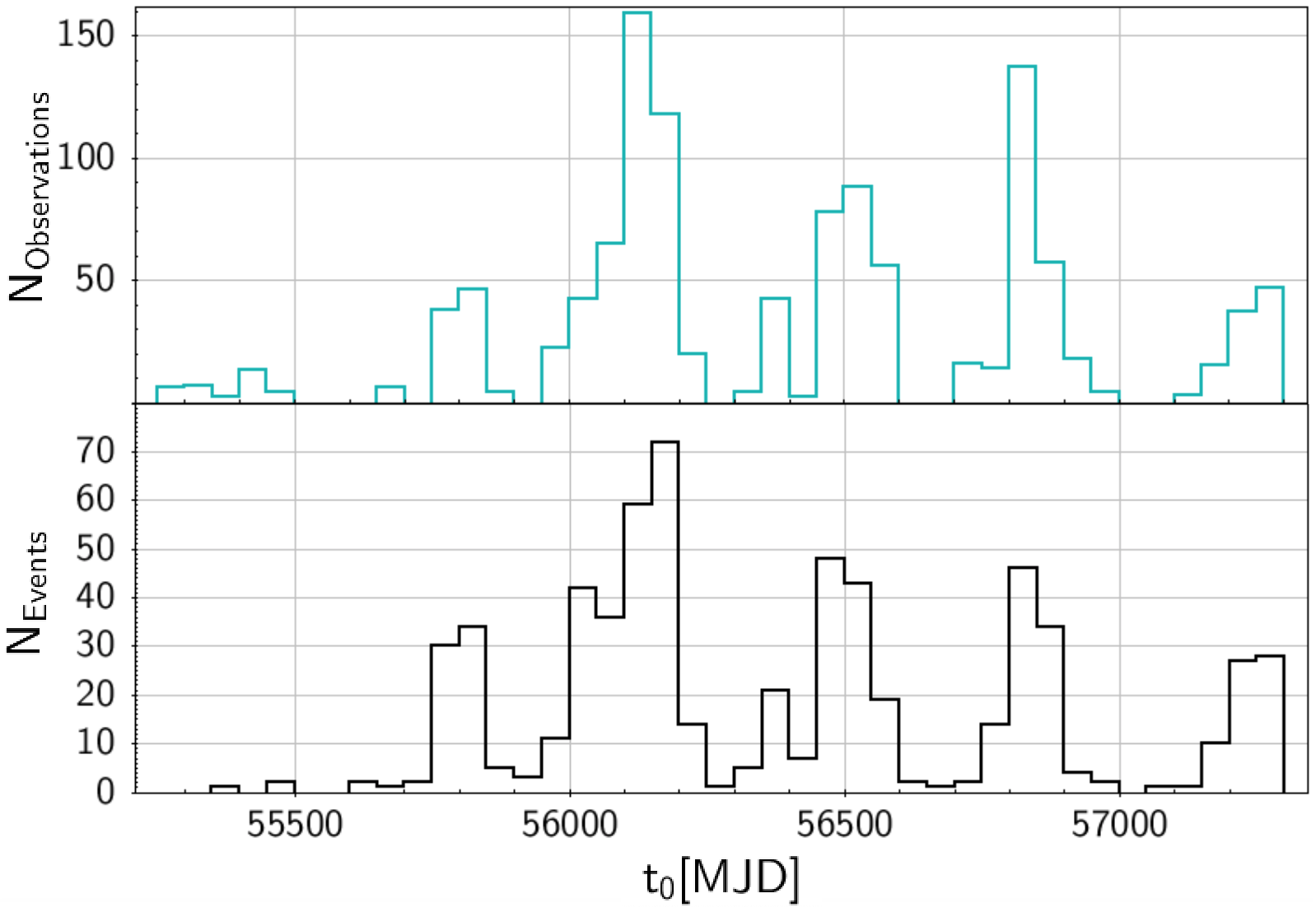}
\caption{Upper panel: Distribution of the number of VVV observations between years 2010 and 2015.
Lower panel: Distogram of the time of maximum magnification ($t_0$) for the detected events between years 2010 and 2015. \\
\label{tmax}}
\end{figure}

The VVV is clearly limited by the sampling efficiency due to the irregular cadence of the observations. 
Some evidences come from the comparison between the histograms of the number of observations of the 14 VVV tiles and the time of maximum magnification ($t_0$) of the first quality sample (Figure~\ref{tmax}).  
The number of observations are shown in the upper panel. 
The serrated distribution is due to yearly cycles of the Bulge observing season.
In general, most yearly observations were taken late in the Bulge season.
Year 2010 was mostly useless because it contained too few epochs of observations and year 2012 was the best year because it contained more epochs than the rest. 
The lower panel shows the number of events detected in the same period of time.
Both histograms follow essentially the same distribution suggesting that the number of the detections is directly affected by the sampling of the VVV. 

\begin{figure}
\epsscale{1.2}
\plotone{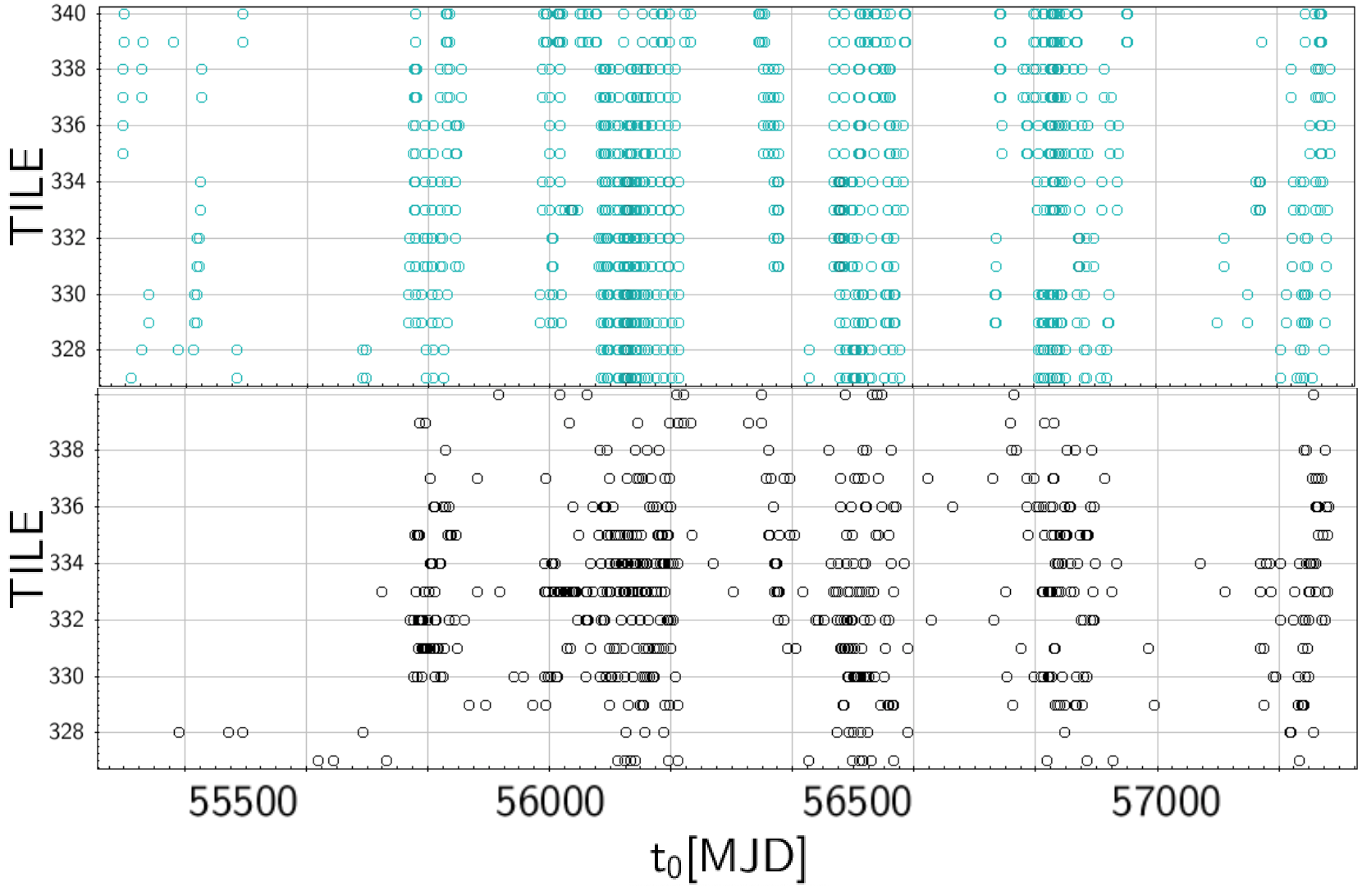}
\caption{Upper panel: Detailed timeline of VVV observations per tile. Each circle represents a single observation of a single tile.
Lower panel: Detailed timeline of microlensing time of maximum magnification ($t_0$) for the events detected per tile. Each circle represents a single event of a single tile. 
The year 2010 was very inefficient for microlensing. \\
\label{obstile}}
\end{figure}

Another clear evidence is the comparison between the timeline of the VVV observations per tile and the time of maximum magnification of the events detected in each tile (Figure~\ref{obstile}). 
This figure illustrates the irregular pattern of observations for our survey, that drives the sampling efficiency. 
We can distinguish five observing seasons where the higher quantity of microlensing events per time were detected.
\textbf{Additionally, the overdensity of events in the innermost tiles ($b332$, $b333$ and $b334$) is also clear.}
The low efficiency present during the first year of the VVV Survey (2010) is evidently due to the small number of observations, and therefore the sampling efficiency will be computed for the complete observational campaign but also for the better epochs (2011-2015).
Although the observations were useful anyway to determine the long timescale behaviour of the light curve baselines.
Again, there is a clear correlation between the observations and the detections. 

\begin{figure}
\epsscale{1.2}
\plotone{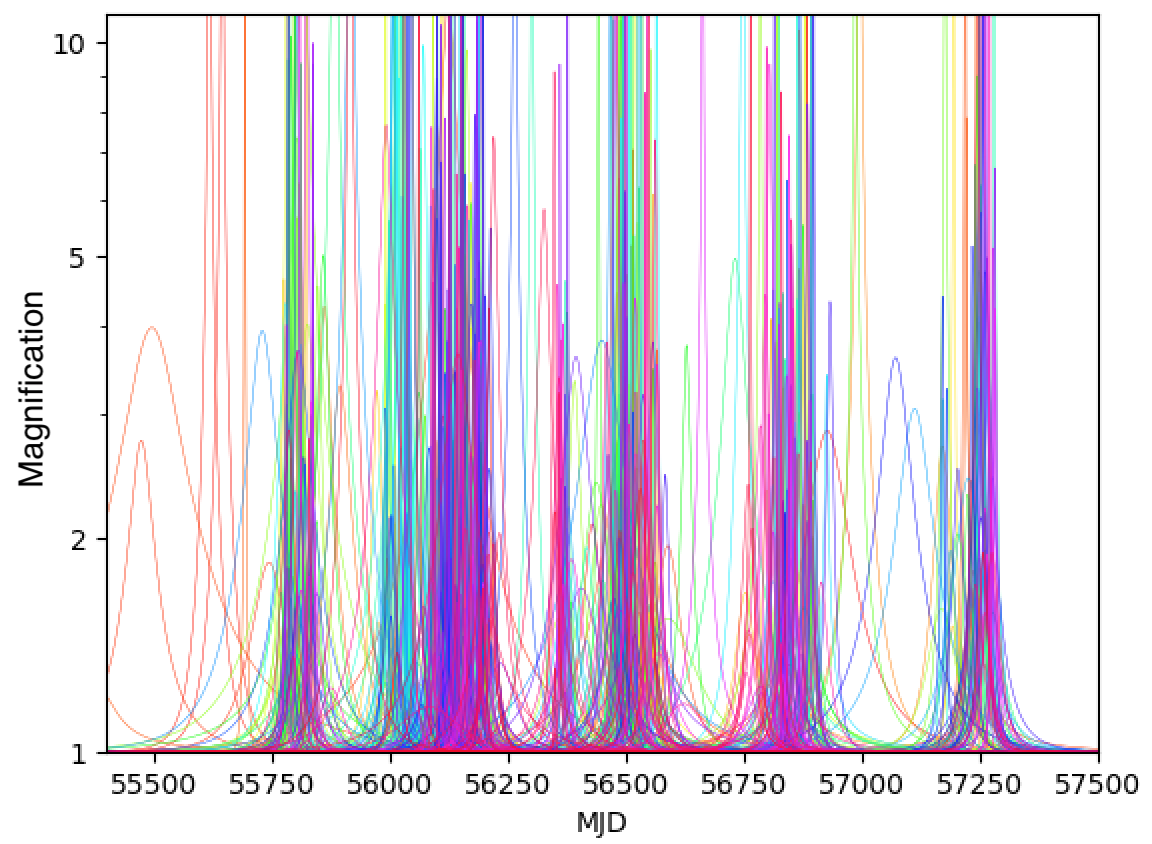}
\caption{Fitted light curves for all VVV microlensing events. 
The long timescale events are more uniformly distributed and do not follow so closely the sampling of the observations, suggesting that these events are more easily detectable.  \\
The animation that shows the microlensing events detected during the operation of the VVV survey is available online. 
In the upper panel a real image of the studied area is shown and each flash correspond to the microlensing events detected in their respective positions. 
The intensity of the flash is associated with the amplitude and the duration with the timescale.
In the lower panel all the light curves of this study are reproduced as a function of time. 
The different colors correspond to the tile of each event and therefore are related to its position.
\label{lc_fits_all}}
\end{figure}

This is also evident if we compute all the light curves together during the five years of observations (Figure~\ref{lc_fits_all}). 
Again, the epochs with more observations show a higher concentration of detections. 
During the months of multiple observations (excluding Jan, Feb, Nov and Dec) we compute an average of 5.5 high magnification events per month.
Additionally, it is clear in which seasons we could expect some long timescale events with a peak at a time with a few or no observations. 
In those few cases the event fulfills all the conditions without any data near the time of highest magnification, but anyway yielding a well constrained light curve. 
This is not possible for short timescale events indicating a stronger timescale dependence in the sampling efficiency. 

\begin{figure}
\epsscale{1.2}
\plotone{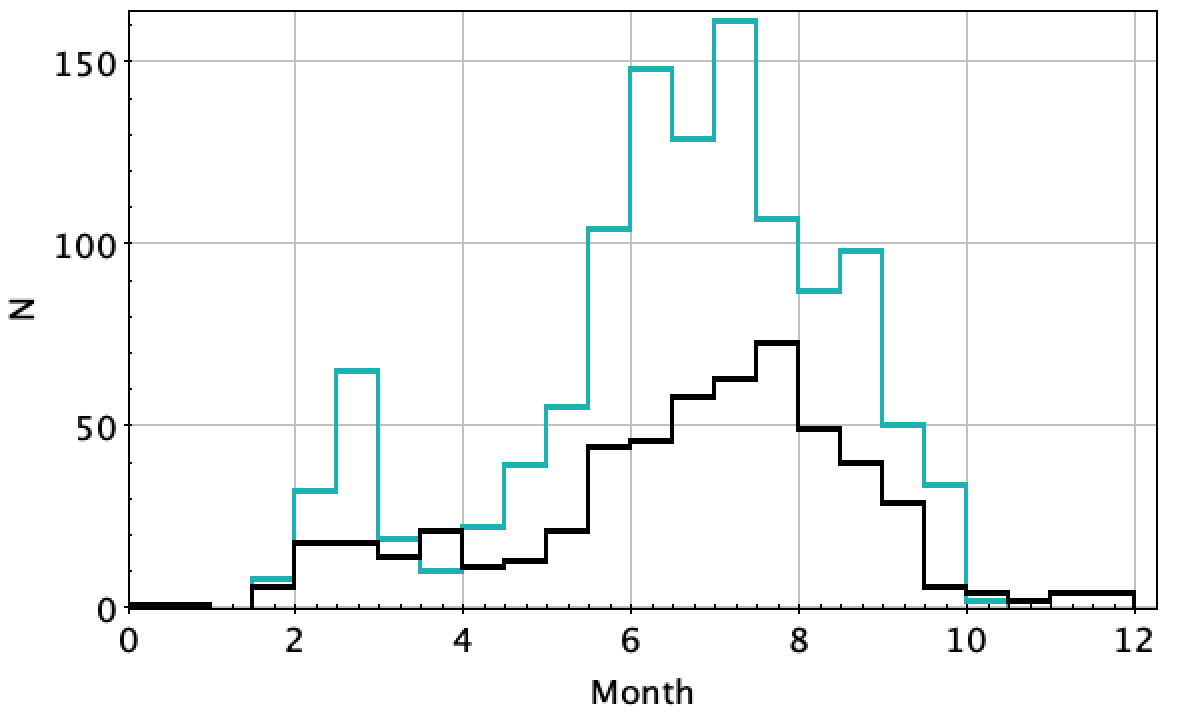}
\caption{Roughly monthly distribution of the number of observations (cyan) compared with the number of detected microlensing events (black).\\
\label{eventdist}}
\end{figure}

Moreover, as the observations for each year were taken during almost the same months, we can also analyze the sampling efficiency monthly. 
Figure~\ref{eventdist} shows the difference between the total observations per month (including the 5 years of survey observations) and the detections per month. 
Both are not uniformly distributed and the months with higher number of events are from April to October, with a bimodality showing a minor peak in April and a major peak in August. 
Again, it is evident that they are directly related. 
%With this evidence, the sampling efficiency should be computed using not only the monthly response but also the yearly response, that is more or less invariant along all the years of observations. 

Additionally,  we considered that the efficiency will depend on the timescale and we expect to be more efficient discovering long timescale events than short timescale events due to the large time coverage but low and irregular cadence of the VVV Survey. 

The objective is to produce multiple fake microlensing light curves with a wide range of parameters ($u_0, t_0, t_E$) and evaluate which ones fulfill all the requirements mentioned in Section \ref{sec:sec2} including the requirements of limiting the magnitude amplitude $<0.1$ in order to be considered a detection. \textbf{The baseline magnitude varies within $10<K_s<17.65$ based on the magnitude distribution of the first quality events (Figure~\ref{KHist_log}). 
The range of values for the 4 parameters ($u_0, t_0, t_E$) is also defined according to the limits and range covered by the first quality sample.}

\textbf{The sampling efficiency was obtained using Monte Carlo simulations to create 1000 fake microlensing events for each timescale, using a grid covering a wide range of timescales ($t_E=$ 1d, 3d, 5d, 10d, 20d, 40d, 60d, 80d, 100d, 150d and 200d). 
We generated $\sim 10^4$ events with a random impact parameter, time of maximum magnification and $K_s$ magnitudes. }
We repeated this procedure until the result converges to one value for the efficiency ($\sim 10$ times), therefore the efficiency for each tile was obtained computing $\sim 10^5$ fake light curves. 

The VVV observational setup is the same throughout the whole survey, which makes this a homogeneous database. 
Even though the light curves may have different observing dates, they all share a similar number of points and limiting magnitudes. 
The slight differences in cadence have been taken into account on a tile by tile basis. 
Therefore this procedure was applied for each tile with slightly different observational campaigns (specifically the central area of each tile as a representative region since the number of observations for each tile remained constant) producing subsequently $\sim 1.5 \times 10^6$ fake microlensing events. 
In this analysis we use a light curve level simulation. The image level simulation is beyond the scope of this paper. 

\textbf{For consistency, as mentioned in Section \ref{sec:sec2}, this procedure was applied to our final sample of 630 events. All of them were recovered as good candidates.}

\begin{figure}
\epsscale{1.2}
\plotone{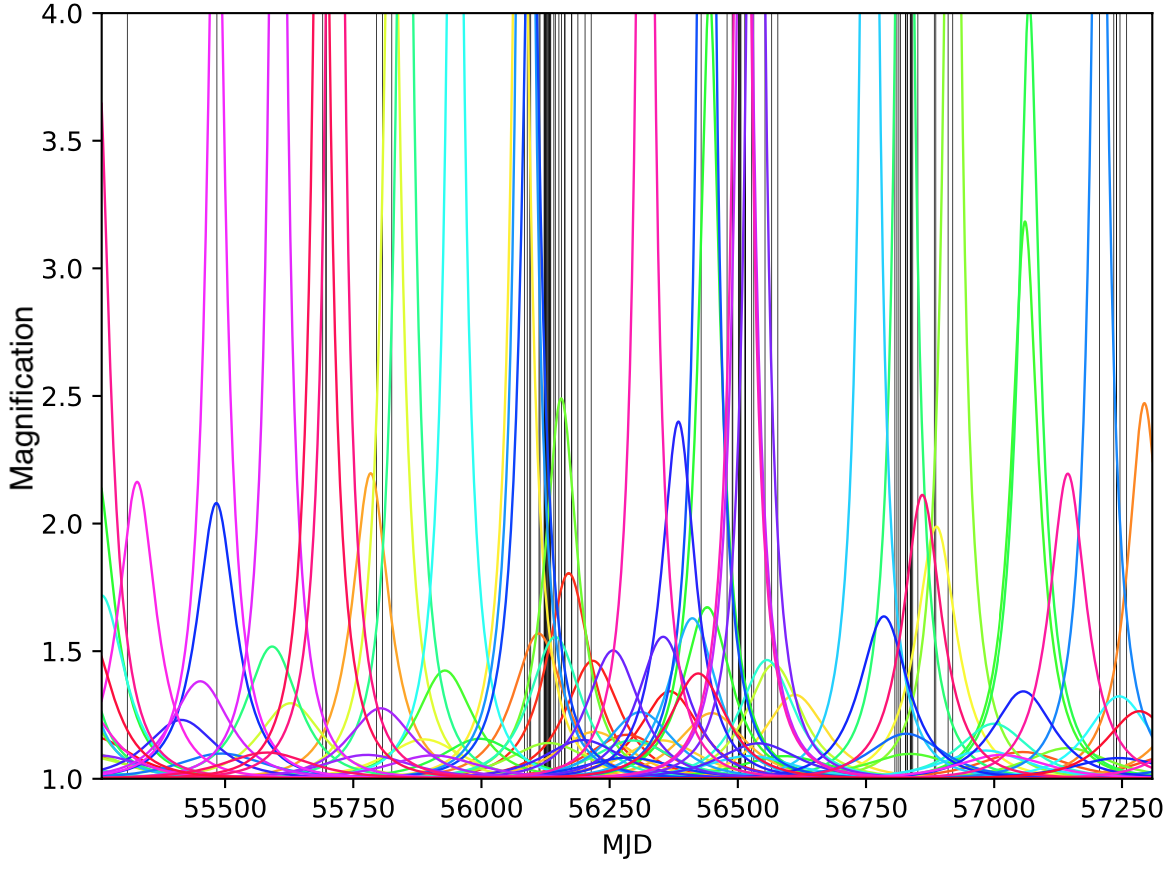}
\caption{Representative plot for the yearly sampling efficiency calculation. 
The 1000 fake events with timescale $t_E = 150$ days are shown as coloured light curves.
The tile colors are ordered from the minimum to maximum longitudes. 
The black vertical lines represent the observations for the tile $b327$.\\
\label{eff}}
\end{figure}

\begin{figure}
\epsscale{1.2}
\plotone{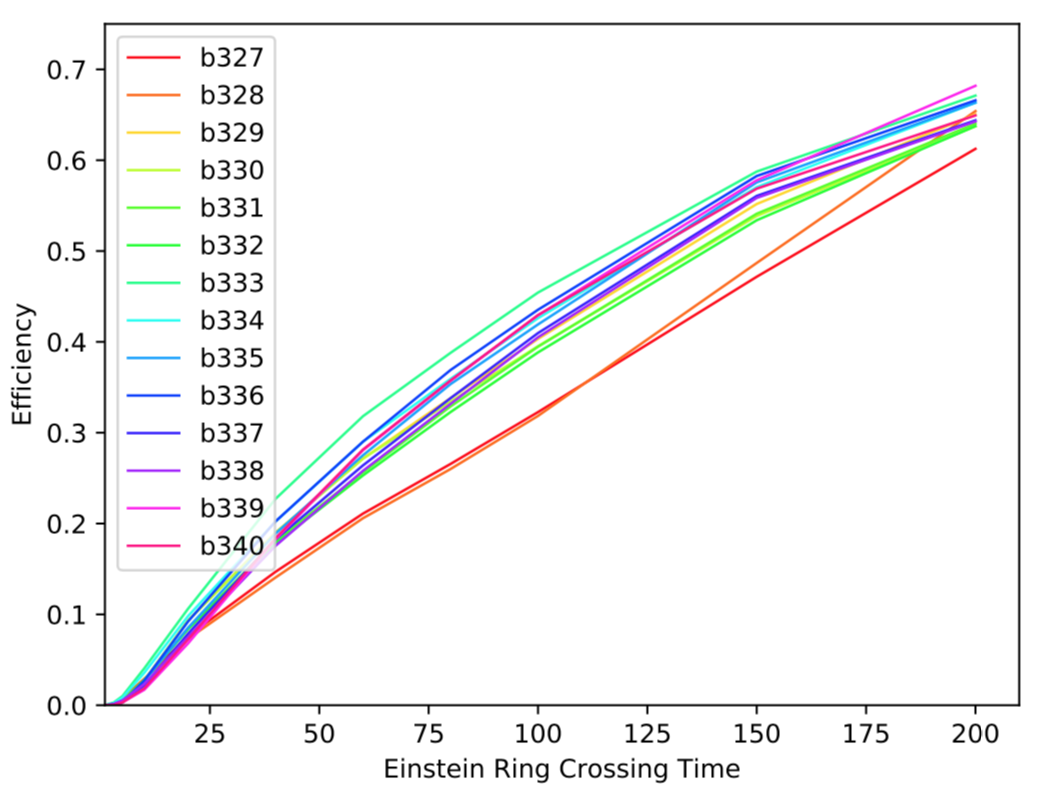}
\caption{Sampling efficiency for the 14 VVV tiles vs Einstein radius crossing time ($t_E$). 
The tiles with lower mean efficiency are the tiles with a lower amount of observations ($b327$ and $b328$)\\
\label{eff2}}
\end{figure}

An example of the calculation of the detection efficiency is shown in Figure~\ref{eff}.
The VVV observations are represented as black vertical lines and the 1000 simulated events are plotted with multiple colors. 
in this case the timescale is fixed ($t_E = 150$ days) with random amplitudes and $t_0$. 
The efficiency computed for every tile and timescale are listed in Table~\ref{tab:table3}, and the respective timescale dependence is shown in Figure~\ref{eff2}. 
The tiles with the lower efficiency correspond to the tiles with fewer observations.
This confirm that the VVV Survey is more efficient discovering long timescale events. \\ 

%We repeated the same procedure for the monthly sampling efficiency.
%The histogram shows the observations per month during the 4 years of observations (2011-2015).  
%The largest number of observations is concentrated on July, and spreads from April to October. 
%The light curves of the complete sample are plotted above the histogram in order to show the concentration of events detected with the maximum magnification during the mentioned months.

%\begin{figure}
%\epsscale{1.2}
%\plotone{obsLClog.png}
%\caption{Roughly monthly distribution of the number of observations (as in the previous figure) overlaying the fitted light curves for all microlensing events.\\
%\label{obsLClog}}
%\end{figure}

%Percentage recovered $38\%$
%minus low amp $\sim 10\%$
% Errors of photometry in k and j ?
%Corrections for foreground disk using the magnitudes and CMD

%%%%%%%%%%%%%%%%%%%%%%%%%%%%%%%%%%%%%%%%%%%%%%%%%%%%%%%%%%%%

\section{Timescale Distribution}
\label{sec:sec8}
The only important physical parameter obtained from the standard microlensing model fitting procedure is the Einstein radius crossing time ($t_E$) also called the microlensing timescale. 
The timescale is related to the mass of the lens, but also depends on the relative distances (distance between the observer and the lens $D_L$ and between the observer and the source $D_S$), and on the transverse velocity.
Thus, although the Einstein radius crossing time is extremely degenerate, the timescale distribution of the sample gives an indication of the masses and velocities of the lenses towards the Galactic Bulge. 
Therefore, the timescale distribution depends on the mass function and the velocity dispersion of the lenses.

The mean timescale for the complete sample corrected by efficiency was computed in \cite{navarro18}. 
The value obtained is $17.4 \pm 1.0$ days for the complete sample and $20.7 \pm 1.0$ days for the RC sources subsample. 
For the model without including the blending parameter the mean timescale is slightly lower $13.9 \pm 1.0$ and $16.5 \pm 1.0 $ days for the RC stars which suggest that the typical lenses are main sequence stars and white dwarfs in both cases. 
Our timescale distribution is also consistent with the findings of \cite{mao96}, who showed that the asymptotic behaviour at long and short timescales follow power-law distributions.

Due to the wide range of longitudes covered by our sample ($-10.00^o \leq l \leq 10.44^o$), the mean timescale is not very representative. 
For this reason, we present the mean timescale distribution for different longitudes in Figure~\ref{tdist}. 
The mean timescale shows a significant change with longitude, decreasing as we approach to the Galactic center (see \cite{navarro17}).  
This behaviour was previously detected by \cite{lukaz15}, but in an area further away from the Galactic center $-2^o \leq b \leq -4$, where the timescale distribution can be affected by different effects and different contributions. 

\begin{figure}
\epsscale{1.2}
\plotone{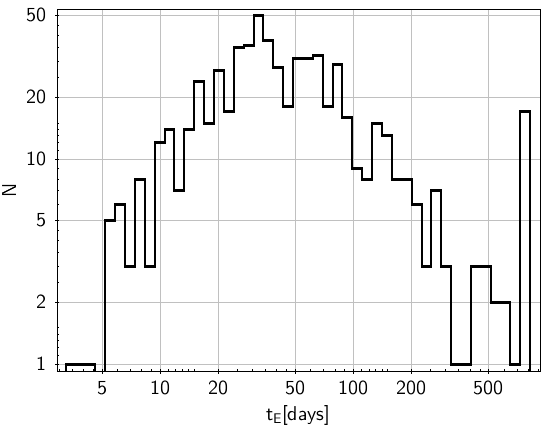}
\caption{Raw timescale distribution of the complete sample. The efficiency corrected timescale distribution is showed in \cite{navarro18}. \\
\label{tdist}}
\end{figure}

%The timescale distribution for the different latitudes is also affected by the efficiency. Thus, we corrected the timescale distribution by the efficiency. The distribution is more affected by the tails of the timescale distribution due to our incompleteness specially in the short timescale tail. 

\begin{figure}
\epsscale{1.2}
\plotone{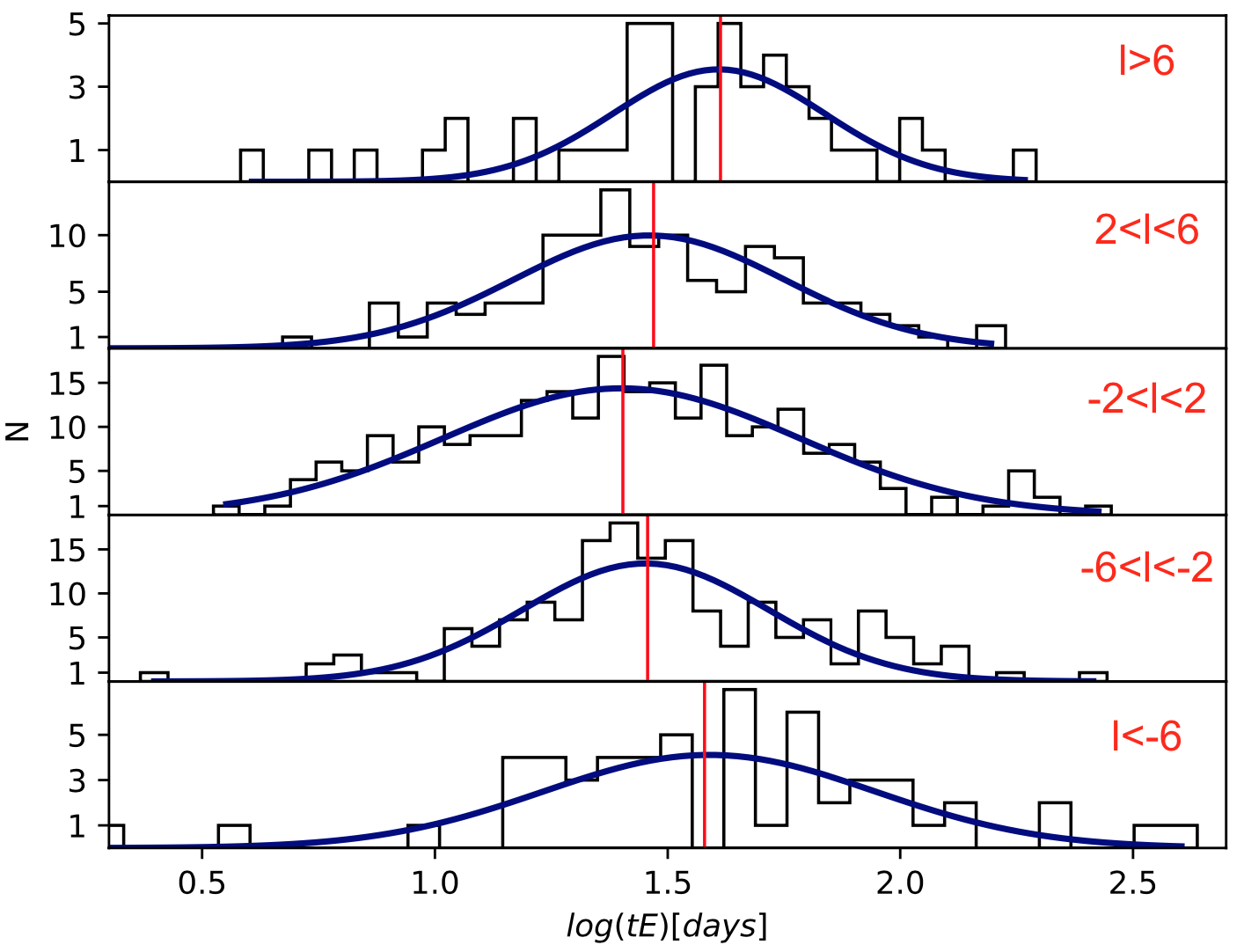}
\caption{Timescale distribution for different ranges of longitudes. 
The longitude range for each histogram is written in the top right corner. 
The red vertical lines show the mean timescale for each histogram.
Is evident the decrease of the mean value for the timescale as we approach to the center from both, positive and negative longitudes. \\
\label{tdist}}
\end{figure}

\section{Proper Motions}
\label{sec:sec9}
The extremely large reddening, and the presence of both disk and Bulge stars along any line of sight toward the Galactic center makes difficult to adopt a membership criterion for the sources, based solely on their position in the CMD shown in Figure~\ref{cmdh} and Figure~\ref{cmdw}.
Proper motions (PM) technique can be a very useful tool to isolate kinematically different stellar populations and help in classifying the sources of microlensing events. 
Unfortunately, observations by the Gaia satellite \citep{gaia18} are severely limited by the dust extinction in this region, and most of our sources are not present in the Gaia database. 
However, the VVV data can penetrate gas and dust in the plane and PMs can be accurately measured in these regions (e.g. \cite{contreras17}). 

\begin{figure}
\epsscale{1.2}
\plotone{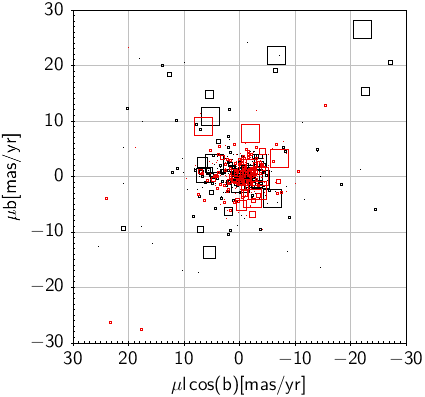}
\caption{Vector Point Diagram of the proper motions for the microlensing events. 
The squares sizes are proportional to the Einstein radius crossing time in a logarithmic scale. 
The black and red squares label the complete sample and Red Clump sources respectively.
The units are millarcsec per year. 
A few high PM objects are clear, and these are probably sources located in the foreground disk. \\
\label{pm}}
\end{figure}

Accordingly, we used the VVV epochs acquired by the survey between 2010 and 2015, to derive the PMs of all the sources. 
To this aim, we employed the same procedure described in detail in \cite{contreras17}, which returns values relative to the mean motion of Bulge Red Giant Branch (RGB) stars. 
The results are plotted in the Vector Point Diagram (VPD) shown in Figure~\ref{pm}, where the PMs of lens sources are weighted by the Einstein radius crossing times (larger squares correspond to larger timescales). 
As expected, the PM distribution is elongated along the longitude axis due to the presence of the Bulge and disk components that partially overlap in the VPD.

As Figure~\ref{pm} shows, the vast majority of events have PMs consistent with sources located in the Galactic Bulge, with mean relative PMs close to zero along both Galactic latitude and longitude. 
These sources exhibit hot kinematics, with low rotation and large velocity dispersion.
 
\begin{figure}
\epsscale{1.2}
\plotone{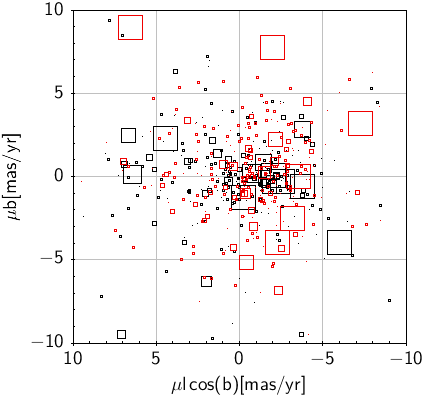}
\caption{Zoom in of the Vector Point Diagram of the proper motions of the microlensing events presented in Figure~\ref{pm}.
The squares sizes are proportional to the Einstein radius crossing time in a logarithmic scale. 
The black and red squares label the complete sample and Red Clump sources respectively.
The units are millarcsec per year. 
 \\
\label{pmzoom}}
\end{figure}

In addition, Figure~\ref{pm} shows a tail of high positive longitude components of the PMs. 
Disk stars are flatter in this plane, due to their smaller velocity dispersion and larger rotation. 
Thus, the sources producing the tail component are consistent with foreground disk sources, located in front of the Milky Way Bulge. 
A caveat is that this analysis is valid for bright sources with little or no blending, therefore we show the VPD also for the red clump sources in red and they follow the same trend.
Figure~\ref{pmzoom} show the innermost area of the VPD of Figure~\ref{pm}

The VVV classification of lens sources based on PMs will improve as the time baseline increases, with the recent extension of the survey called VVVX, that will map these fields until year 2020.

%Many of the largest timescale events have higher PMs, as expected if they are relatively nearby disk-disk events. However,

%%%%%%%%%%%%%%%%%%%%%%%%%%%%%%%%%%%%%%%%%%%%%%%%%%%%%%%%%%%%

\section{Microlensing with WFIRST}
\label{sec:sec10}
The future microlensing plans with WFIRST are described by \citealt{Green12}, \citealt{Spergel15} and \citealt{penny18}.
In light of the present results with the VVV, the detection of so many events by a survey that was not especially designed for microlensing opens the possibility to do a much more efficient survey with WFIRST, if there is a $K$-band filter. Note that \cite{Stauffer18} argued strongly for the need of a $K$-band filter in WFIRST for many other important scientific considerations. 
Based on our results, we can crudely estimate the yield of a microlensing experiment with WFIRST that covers the same area mapped here (20.7 sq. deg. at $l=0^\circ$). 
Taking $2 \times 20$ sec exposures to reach $K\sim 21$ mag, WFIRST can cover the equivalent area in approximately 6 hours (computation based on a talk by R. Benjamin). 
Therefore, WFIRST would be extremely efficient. 
In just a two-month campaign it could detect $>10000$ microlensing events with timescales $1<t_E<60$ days, allowing studies of:

\begin {enumerate}
\item planetary microlensing and free-floating planet searches
\item binary microlensing events
\item finite source effects
\item Galactic structure, including events from the far-side of the Milky Way
\item microlensing by the central supermassive Black Hole
\end {enumerate}

In addition, a continuing monitoring of the area at lower cadence requiring only a couple of dozen extra days of WFIRST observations would allow to study:

\begin {enumerate}
\item parallax and xallarap events (in combination with a dedicated ground-based observation campaign)
\item massive objects (Black Holes, Neutron Stars)
\end {enumerate}

This would indeed be an independent way of budgeting the contribution of massive remnants in the Galactic centre region. 
We most strongly advocate a microlensing survey in this whole area with WFIRST.

%%%%%%%%%%%%%%%%%%%%%%%%%%%%%%%%%%%%%%%%%%%%%%%%%%%%%%%%%%%%

\section{Comparison with other microlensing experiments}
\label{sec:sec11}
Even though they are not directly comparable quantitatively, because they explore different areas and because the different observational strategies applied on every case may borne out peculiar results. 
We would like to stress the similarities and differences with some of the other optical and near-IR microlensing experiments. 
Therefore, a qualitative comparison with the result obtained with previous microlensing surveys is considered. 

Compared with MACHO, we have similar number of epochs cover a similar total area as their first Bulge seasons. 

Compared with OGLE in its various incarnations, we have much fewer observations. 
The huge relative advantage of OGLE is then not only in a better definition of the microlensing event but also of the baseline. 
The higher cadence of OGLE allows to detect more events in the short timescale regime, thus the observational data is more complete, otherwise since the observations are done in optical bandpasses is not possible to reach low latitudes. \\

From this work we find a relatively low sampling efficiency compared with more sophisticated and dedicated microlensing experiments carried out in the optical like MACHO, OGLE, MOA and EROS. 
In particular OGLE is the longest lasting experiment and our sampling efficiency should clearly be worse in comparison due to the small number of epochs (70-100 compared with thousands) and the small number of years of observations (5 compared with $>25$).

The photometric efficiencies are not directly comparable, i.e. the completeness maps are completely different for each survey, but the fact that we are observing in the near-IR bandpasses allows us to sample the Bulge RC very well and to detect events in the most crowded and reddened low-latitude areas, specifically for the region studied here, along the Galactic plane, where these optical surveys are blind.

 %From this work we find a relatively low efficiency compared with more sophisticated and dedicated microlensing experiments carried out in the optical like MACHO, OGLE, MOA and EROS. In particular OGLE is the longest lasting experiment and our efficiency should clearly be worse in comparison. The photometric efficiencies are comparable  (albeit at different wavelengths), but our sampling efficiency is low, due to the small number of epochs (70-100 compared with thousands) and the small number of years of observations (5 compared with $>25$).
 
Compared with the UKIRT microlensing survey, they are very similar in many respects: 
they have a similar near-IR search in the $H$-band, with similar limiting magnitudes, at a similar size telescope with a similar wide-field camera. 
Also, we observed from 2010 to 2015, and they are observing since 2015. 
We cover a much more extended area, but they have better cadence favouring the detection of short timescale events \citep{Shvartzvald17}.
The continuation of the UKIRT microlensing search is very important, and based on the present results, we should be very optimistic about the upcoming results from such experiment.

The observational data are sensitive to many Galactic parameters. 
The only way to associate them is through models which include parameters such as stellar velocities and spatial distributions, inicial mass function, etc., that are poorly constrained observationally, which may allow several ways to match microlensing data with very different Galactic models. 

%The observational data are sensitive to many Galactic parameters. The only way to associate them is through models which include parameters such as stellar velocities and spatial distributions, inicial mass function, etc., that are poorly constrained. 
%Since they are related to the Galaxy through models with assumed values of the parameters. Most of the Galactic parameters such as stellar velocities as a function of position are poorly constrained observationally, which may allow several ways to match microlensing data with very different Galactic models. 

Given this uncertainty, we wish to investigate the sensitivity of timescale frequency distributions and optical depths to changes in the values of several Galactic parameters, and for example, to compare the model microlensing measures at every step with the empirical measures deduced from the microlensing events followed by the MACHO group during the summer of 1993 (Alcock et al. 1996). 
Such models are clearly needed, but beyond the scope of the present study.

  %Bright sources - sufficient to estimates detection efficiency based on the sampling of the lc alones

%%%%%%%%%%%%%%%%%%%%%%%%%%%%%%%%%%%%%%%%%%%%%%%%%%%%%%%%%%%%

\section{Conclusions}
\label{sec:sec12}
We present the detailed description of the searching procedure and the analysis of the 630 new microlensing events (290 located in the red clump of the CMD) detected in low latitudes of the Galaxy. 
The area of $20.68 deg^2$ located within $-10.00^o \leq l \leq 10.44^o$ and $ -0.46^o \leq b \leq 0.65^o$ was observed by the VVV Survey in the near-IR. 

We present the table of the first quality events along with the Galactic coordinates, $J-K_s$ colors, $K_s$ magnitudes and the parameters obtained by the standard microlensing model. 
Additionally, we present the Color magnitude diagram ($K_s$ vs $H-K_s$) and proper motions of the final sample. 
The timescale distribution is analysed for different longitudes showing a decrease in the mean timescale as we approach to the Galactic center.
For sake of completeness we present the second quality events along with the $K_s$ magnitudes and the Galactic coordinates. 
The detailed analysis of events that do not follow the standard microlensing model (binaries, parallax effect, etc...) is beyond the scope of this paper and is proposed for the future. 
We study the efficiency (photometric and sampling) of the survey and find that de VVV is a powerful tool to detect long timescale events. 

To conclude,  the VVV Survey is a powerful tool to detect microlensing events and to study this population at low latitudes where the extinction and crowding restrain the observations in optical bandpasses. 
This can be useful to plan the WFIRST microlensing observation campaigns  (\citealt{Green12}, \citealt{Spergel15}) and Euclid Survey \citep{Hamolli15}.
We also proposed an efficient survey of microlensing events in the central Galactic plane that would discover $>10^4$ events, enabling studies of extrasolar planets and free-floating planets, Galactic structure, finite source effects, parallax events, and search for massive BHs.

\acknowledgments
We gratefully acknowledge data from the ESO Public Survey program ID 179.B-2002 taken with the VISTA telescope, and products from the Cambridge Astronomical Survey Unit (CASU).

\newpage

\section*{Appendix}

%%%%%%%%%%%%%%%%%%%%%   TABLES %%%%%%%%%%%%%%%%%%%%%%%%%%%%%%%

\begin{longtable}{l l l}
\tablecaption{Number of light curves analyzed in the different steps of the selection procedure.} 
\label{tab:table0} \\
\tablehead{
\colhead{Step description} & \colhead{Number of light curves} & \colhead{Number of light curves analyzed} \\
  & \colhead{obtained per tile} & \colhead{obtained in total} 
}
Complete sample of light curves & $\sim 4-5 \times 10^6$ & $\sim 10^7$  \\
Number of data points of each light curve $>20$ & $\sim 2-4 \times 10^6$ & $\sim 5 \times 10^6$  \\
Fitting procedure; No binaries,  & $2500-5000$ & $\sim 5 \times 10^4$ \\
$t_E <$500 days, magnitude amplitude $>0.1$, quality index $<0.002$ & & \\
Visual inspection & $13-119$ & 655  \\
 \end{longtable}

\begin{longtable*}{ l l l l l l l l l l l}
\tablecaption{VVV Survey first quality microlensing events data with their respective positions in Galactic coordinates, baseline $K_s$ magnitude, Color and the parameters obtained using the standard microlensing model including the blending ($f_{bl}$). 
} \label{tab:table1} \\
\tablehead{
\colhead{Tile} & \colhead{ID} & \colhead{l} & \colhead{b} & \colhead{$K_s$} & \colhead{$J-K$} & \colhead{Amp} & \colhead{$u_{0}$} & \colhead{$t_{0}$} & \colhead{$t_E$} & \colhead{$f_{bl}$}  \\
 & &  {\tiny (deg)} &  {\tiny (deg)} & {\tiny (mag)} & {\tiny (mag)}  & {\tiny (mag)} &  & {\tiny (MJD)} & {\tiny (days)}  & \\ }
327 & 23349 & -9.50040 & 0.23641 & 16.67 & $>$4.79 & 1.31 & 0.37 $\pm$ 0.38 & 56926.82 $\pm$ 24.01 & 222.08 $\pm$ 135.29 & 1.00 $\pm$ 1.48 \\
327 & 60661 & -9.10190 & 0.15904 & 12.11 & 2.22 & 0.81 & 0.52 $\pm$ 0.45 & 56427.23 $\pm$ 11.76 & 77.73 $\pm$ 30.61 & 1.00 $\pm$ 1.36 \\
327 & 12866 & -8.85924 & 0.43688 & 14.99 & 4.14 & 1.42 & 0.11 $\pm$ 0.06 & 56210.38 $\pm$ 0.71 & 47.72 $\pm$ 10.44 & 0.42 $\pm$ 0.19 \\
327 & 27925 & -8.57640 & 0.40466 & 14.67 & 4.93 & 0.78 & 0.49 $\pm$ 0.15 & 56497.41 $\pm$ 0.14 & 15.50 $\pm$ 2.91 & 1.00 $\pm$ 0.45 \\
327 & 26527 & -8.87632 & 0.54326 & 14.99 & 4.02 & 0.50 & 0.18 $\pm$ 0.10 & 56529.63 $\pm$ 1.12 & 139.31 $\pm$ 62.24 & 0.12 $\pm$ 0.08 \\
327 & 77677 & -9.91889 & -0.05814 & 14.64 & 1.05 & 2.12 & 0.01 $\pm$ 0.00 & 56515.21 $\pm$ 0.05 & 108.00 $\pm$ 35.02 & 0.06 $\pm$ 0.02 \\
327 & 81144 & -9.93961 & -0.06510 & 12.16 & 2.64 & 0.81 & 0.00  $\pm >$ 2  &55645.26 $\pm$ 6.87 & 86.41 $\pm$ 15.07 & 1.00 $\pm$ 0.50 \\
327 & 9538 & -9.51913 & 0.62890 & 14.41 & 2.81 & 0.29 & 1.14  $\pm >$ 2  &56509.41 $\pm$ 0.86 & 16.75 $\pm >$ 100  &1.00 $\pm$ 7.26 \\
327 & 21550 & -8.74355 & -0.12582 & 14.78 & 1.12 & 0.98 & 0.00 $\pm$ 0.01 & 56134.11 $\pm$ 0.05 & 150.04 $\pm >$ 100  &0.00 $\pm$ 0.01 \\
327 & 6927 & -9.26352 & -0.09862 & 15.36 & 2.90 & 0.47 & 0.00  $\pm >$ 2  &55618.93 $\pm$ 13.67 & 85.78 $\pm$ 54.49 & 1.00 $\pm$ 1.55 \\
327 & 2328 & -8.66127 & 0.63808 & 14.41 & 4.95 & 0.39 & 0.87 $\pm$ 1.22 & 56565.99 $\pm$ 2.12 & 32.15 $\pm$ 27.51 & 1.00 $\pm$ 2.71 \\
327 & 27976 & -8.68417 & 0.53720 & 16.20 & 3.36 & 1.44 & 0.27 $\pm$ 0.11 & 56818.82 $\pm$ 0.33 & 42.32 $\pm$ 12.90 & 1.00 $\pm$ 0.52 \\
327 & 4715 & -8.64538 & 0.09324 & 15.87 & 3.59 & 2.79 & 0.00  $\pm >$ 2  &56884.26 $\pm$ 2.04 & 22.75 $\pm >$ 100  &0.59 $\pm$ 3.02 \\
327 & 30934 & -8.59898 & 0.03953 & 13.86 & 2.62 & 0.33 & 0.60 $\pm$ 0.35 & 56196.06 $\pm$ 1.92 & 62.93 $\pm$ 22.09 & 0.43 $\pm$ 0.40 \\
327 & 20462 & -9.41172 & -0.32384 & 14.86 & $>$6.23 & 1.18 & 0.35 $\pm$ 0.08 & 56112.84 $\pm$ 0.30 & 57.54 $\pm$ 8.91 & 1.00 $\pm$ 0.29 \\
327 & 7903 & -8.54318 & 0.27034 & 14.59 & 3.23 & 1.14 & 0.16 $\pm$ 0.25 & 56141.46 $\pm$ 1.98 & 85.98 $\pm >$ 100  &0.36 $\pm$ 0.63 \\
327 & 19181 & -8.72745 & 0.06383 & 16.13 & $>$5.25 & 0.67 & 0.62 $\pm$ 1.16 & 55732.79 $\pm$ 18.06 & 250.61 $\pm >$ 100  &1.00 $\pm$ 3.07 \\
327 & 50494 & -9.80524 & -0.19031 & 16.66 & 2.02 & 2.73 & 0.10 $\pm$ 0.43 & 56124.72 $\pm$ 6.82 & 293.34 $\pm >$ 100  &1.00 $\pm$ 4.58 \\
327 & 51715 & -8.76551 & 0.17019 & 12.50 & 3.36 & 0.32 & 0.22 $\pm$ 0.60 & 57233.43 $\pm$ 2.00 & 72.46 $\pm >$ 100  &0.09 $\pm$ 0.30 \\
328 & 21982 & -7.15966 & -0.31066 & 14.66 & 2.69 & 1.10 & 0.16 $\pm$ 0.16 & 55472.23 $\pm$ 11.91 & 254.17 $\pm$ 134.02 & 0.34 $\pm$ 0.35 \\
328 & 38942 & -7.13369 & 0.52860 & 15.65 & $>$5.60 & 0.66 & 0.66 $\pm$ 0.75 & 56125.75 $\pm$ 1.44 & 54.27 $\pm$ 42.40 & 1.00 $\pm$ 1.95 \\
328 & 72058 & -7.43431 & -0.25762 & 17.05 & $>$4.41 & 2.01 & 0.00  $\pm >$ 2  &56157.06 $\pm$ 0.81 & 9.85 $\pm >$ 100  &0.63 $\pm$ 2.01 \\
328 & 49552 & -7.45715 & 0.35790 & 13.13 & 3.68 & 0.31 & 0.99  $\pm >$ 2  &56847.68 $\pm$ 5.19 & 59.38 $\pm >$ 100  &0.95 $\pm$ 10.23 \\
328 & 34307 & -7.17794 & 0.19524 & 14.31 & $>$6.74 & 0.31 & 0.04 $\pm$ 0.04 & 57256.64 $\pm$ 3.99 & 750.00 $\pm$ 522.78 & 0.02 $\pm$ 0.01 \\
328 & 15560 & -7.17864 & -0.11907 & 15.15 & 2.68 & 2.18 & 0.00  $\pm >$ 2  &56125.57 $\pm$ 0.01 & 3.86 $\pm$ 1.40 & 1.00 $\pm$ 0.58 \\
328 & 51742 & -7.15474 & -0.19830 & 13.46 & 3.16 & 0.70 & 0.57 $\pm$ 0.69 & 56589.44 $\pm$ 15.33 & 93.79 $\pm$ 62.77 & 1.00 $\pm$ 2.05 \\
328 & 81246 & -7.20060 & -0.26304 & 15.14 & 2.81 & 1.51 & 0.20 $\pm$ 0.06 & 56511.61 $\pm$ 0.12 & 14.04 $\pm$ 2.36 & 1.00 $\pm$ 0.29 \\
328 & 65766 & -7.39676 & -0.22788 & nan & nan & nan & 0.30 $\pm$ 0.28 & 55494.68 $\pm$ 14.22 & 347.45 $\pm$ 226.77 & 1.00 $\pm$ 1.22 \\
328 & 38433 & -8.31739 & 0.36769 & 16.32 & 4.12 & 1.34 & 0.31 $\pm$ 0.21 & 56473.63 $\pm$ 12.82 & 295.94 $\pm$ 120.60 & 1.00 $\pm$ 0.91 \\
328 & 28493 & -7.71797 & 0.03549 & 14.52 & 4.79 & 0.57 & 0.69 $\pm$ 0.35 & 56188.67 $\pm$ 0.72 & 28.24 $\pm$ 8.90 & 1.00 $\pm$ 0.88 \\
328 & 51701 & -7.65567 & -0.01853 & 15.27 & $>$5.92 & 0.88 & 0.49 $\pm$ 0.21 & 56500.04 $\pm$ 0.85 & 78.47 $\pm$ 22.84 & 1.00 $\pm$ 0.65 \\
328 & 47319 & -7.38468 & 0.18311 & 17.03 & $>$4.39 & 1.98 & 0.05 $\pm$ 0.05 & 56493.47 $\pm$ 0.32 & 34.29 $\pm$ 14.78 & 0.62 $\pm$ 0.36 \\
328 & 39497 & -8.13846 & -0.18339 & 16.83 & $>$4.58 & 1.08 & 0.00  $\pm >$ 2  &57218.73 $\pm$ 1.37 & 32.71 $\pm$ 30.22 & 1.00 $\pm$ 1.86 \\
328 & 48714 & -8.51069 & -0.00228 & 17.11 & 2.48 & 3.37 & 0.04 $\pm$ 0.01 & 56156.53 $\pm$ 0.19 & 21.53 $\pm$ 5.18 & 1.00 $\pm$ 0.33 \\
328 & 4792 & -8.48756 & -0.27652 & 12.94 & 2.93 & 0.22 & 0.10 $\pm$ 0.08 & 55389.94 $\pm$ 10.28 & 586.64 $\pm$ 341.10 & 0.02 $\pm$ 0.02 \\
328 & 47538 & -7.28364 & 0.35930 & 14.87 & $>$6.28 & 0.96 & 0.02 $\pm$ 0.01 & 56523.12 $\pm$ 0.43 & 490.43 $\pm$ 210.66 & 0.04 $\pm$ 0.02 \\
328 & 83166-1746 & -8.34604 & 0.09827 & 15.24 & 3.85 & 0.78 & 0.06 $\pm$ 0.43 & 57221.15 $\pm$ 0.44 & 44.96 $\pm$ 35.34 & 0.41 $\pm$ 0.54 \\
... &  &  &  &  &  & & &  &  &  \\
\end{longtable*}
\footnotemark{This table will be available in its entirety in machine-readable form.} \\
\footnotemark{One standard deviation errors are presented along each parameter obtained from the microlensing model fitting procedure} \\
\indent \footnotemark{Typical positional errors are 0.1 arcsec \citep{Smith17}. } \\
\indent \footnotemark{Typical photometric errors are $\sigma_{K_s} = 0.01$ mag, and $\sigma_{J, H} = 0.03$ mag \citep{saito12, contreras17, alonso18}. } \\

\begin{longtable}{l l l l l}
\tablecaption{VVV Survey second quality microlensing events data with their respective positions in Galactic coordinates, baseline $K_s$ magnitude.} \label{tab:table2} \\
\tablehead{
\colhead{Tile} & \colhead{ID} & \colhead{l} & \colhead{b} & \colhead{$K_s$}  \\
 & & {\tiny (deg)}  & {\tiny (deg)} & {\tiny (mag)}  \\ }
327 & 44776 & -9.85935 & -0.38927 & 12.29 \\
327 & 36359 & -9.56105 & -0.36532 & 14.09 \\
327 & 42538 & -9.43791 & 0.55426 & 13.11 \\
327 & 8202 & -8.79797 & -0.28197 & 14.36 \\
327 & 8935 & -8.79612 & -0.28367 & 15.75 \\
327 & 44952 & -8.88308 & -0.36659 & 13.96 \\
327 & 45728 & -9.14494 & 0.35358 & 12.79 \\
327 & 15205 & -9.92297 & -0.11414 & 12.71 \\
327 & 55460 & -9.93087 & -0.20375 & 11.86 \\
327 & 51906 & -9.60936 & 0.35375 & 17.41 \\
327 & 2889 & -9.48211 & -0.08854 & 17.35 \\
327 & 7146 & -9.52606 & -0.09738 & 13.94 \\
327 & 7849 & -9.52702 & -0.09884 & 13.55 \\
327 & 8572 & -9.52575 & -0.10030 & 15.16 \\
327 & 9753 & -9.52742 & -0.10273 & 15.66 \\
327 & 17827 & -9.57760 & -0.11904 & 13.84 \\
327 & 1931 & -9.75869 & 0.64509 & 15.05 \\
327 & 67124 & -9.62085 & 0.47951 & 15.52 \\
327 & 61344 & -9.37651 & -0.04158 & 13.58 \\
327 & 72421 & -9.80824 & -0.04772 & 16.75 \\
327 & 10018 & -9.09113 & -0.10203 & 15.15 \\
327 & 11322 & -9.09235 & -0.10476 & 14.72 \\
327 & 12480 & -9.09238 & -0.10709 & 14.54 \\
327 & 63547 & -9.08415 & -0.22403 & 12.61 \\
327 & 41620 & -9.43792 & 0.55427 & 13.11 \\
327 & 82384 & -9.52769 & 0.46225 & 14.04 \\
327 & 81939 & -8.70800 & -0.26137 & 15.71 \\
327 & 82074 & -8.70795 & -0.26169 & 15.84 \\
327 & 3292 & -9.47431 & 0.09514 & 17.27 \\
327 & 78825 & -9.47350 & -0.06291 & 12.99 \\
327 & 8400 & -8.97538 & -0.09867 & 14.95 \\
327 & 24200 & -9.35747 & 0.40805 & 14.44 \\
327 & 36371 & -8.92269 & 0.02898 & 12.60 \\
327 & 48177 & -8.90979 & 0.00547 & 17.10 \\
327 & 72904 & -8.97389 & -0.04364 & 16.10 \\
327 & 30083 & -8.53334 & 0.04144 & 14.33 \\
327 & 37236 & -8.60059 & 0.02698 & 14.21 \\
327 & 92033 & -8.55863 & -0.08926 & 14.30 \\
327 & 4289 & -9.03403 & 0.64183 & 14.83 \\
327 & 36407 & -8.94349 & 0.54389 & 13.53 \\
327 & 7320 & -9.07541 & 0.27092 & 11.61 \\
327 & 88621 & -9.04659 & 0.09834 & 12.60 \\
327 & 71710 & -9.73790 & 0.13494 & 13.58 \\
327 & 76892 & -8.51841 & 0.12615 & 15.85 \\
327 & 3037 & -8.82103 & 0.09699 & 15.28 \\
327 & 42878 & -8.85320 & 0.01529 & 14.66 \\
327 & 44221 & -8.70271 & 0.01302 & 15.14 \\
327 & 58744 & -9.63049 & -0.20990 & 14.57 \\
327 & 7246 & -9.76736 & -0.28329 & 13.75 \\
327 & 46489 & -8.78125 & 0.18179 & 11.70 \\
327 & 83413 & -8.76071 & 0.10369 & 11.72 \\
328 & 83820 & -7.07437 & -0.44394 & 16.65 \\
328 & 53473 & -7.24635 & -0.02886 & 14.54 \\
328 & 3916 & -7.12700 & 0.63964 & 13.85 \\
328 & 1972 & -7.47843 & 0.09815 & 17.11 \\
328 & 53678 & -7.43061 & -0.01978 & 15.93 \\
328 & 36014 & -7.42562 & -0.16508 & 17.38 \\
328 & 75870 & -7.50794 & -0.26678 & 12.96 \\
328 & 75455 & -7.57421 & 0.30196 & 17.30 \\
328 & 46922 & -8.23199 & 0.17859 & 17.01 \\
328 & 58617 & -8.26415 & 0.15215 & 16.09 \\
328 & 74922 & -8.22929 & 0.11654 & 17.00 \\
... &  &  &  &  \\
 \end{longtable}
\footnotemark{This table will be available in its entirety in machine-readable form.} \\
\footnotemark{Typical positional errors are 0.1 arcsec \citep{Smith17}. } \\
\indent \footnotemark{Typical photometric errors are $\sigma_{K_s} = 0.01$ mag, and $\sigma_{J, H} = 0.03$ mag \citep{saito12, contreras17, alonso18}. } \\

 \newpage

\begin{longtable*}{l l l l l l l l l l l l}
\tablecaption{Sampling efficiency for each tile and timescale obtained with the Monte Carlo simulation.} 
\label{tab:table3} \\
\tablehead{
 \colhead{Tile} & \colhead{$t_E = 1$} & \colhead{$t_E = 3$} & \colhead{$t_E = 5$} & \colhead{$t_E = 10$} & \colhead{$t_E = 20$} & \colhead{$t_E = 40$} & \colhead{$t_E = 60$} & \colhead{$t_E = 80$} & \colhead{$t_E = 100$} & \colhead{$t_E = 150$} & \colhead{$t_E = 200$}\\
 &{\tiny (days)} &{\tiny (days)} & {\tiny (days)}&{\tiny (days)}&{\tiny (days)} & {\tiny (days)}& {\tiny (days)} &{\tiny (days)} & {\tiny (days)}  & {\tiny (days) } & {\tiny (days)}
}
$b327$ & 0.0000 & 0.0008 & 0.0075 & 0.0288 & 0.0760 & 0.1469 & 0.2110 & 0.2656 & 0.3227 & 0.4717 & 0.6125 \\
$b328$ & 0.0000 & 0.0008 & 0.0067 & 0.0291 & 0.0731 & 0.1406 & 0.2059 & 0.2600 & 0.3185 & 0.4869 & 0.6540 \\
$b329$ & 0.0000 & 0.0005 & 0.0029 & 0.0272 & 0.0932 & 0.1902 & 0.2716 & 0.3381 & 0.4033 & 0.5518 & 0.6505 \\
$b330$ & 0.0000 & 0.0005 & 0.0029 & 0.0272 & 0.0932 & 0.1897 & 0.2715 & 0.3342 & 0.3952 & 0.5386 & 0.6374 \\
$b331$ & 0.0000 & 0.0017 & 0.0047 & 0.0278 & 0.0839 & 0.1808 & 0.2564 & 0.3289 & 0.3947 & 0.5411 & 0.6402 \\
$b332$ & 0.0000 & 0.0017 & 0.0047 & 0.0278 & 0.0838 & 0.1781 & 0.2530 & 0.3227 & 0.3884 & 0.5339 & 0.6375 \\
$b333$ & 0.0000 & 0.0030 & 0.0100 & 0.0406 & 0.1062 & 0.2272 & 0.3182 & 0.3879 & 0.4543 & 0.5875 & 0.6711 \\
$b334$ & 0.0000 & 0.0023 & 0.0084 & 0.0357 & 0.0981 & 0.2022 & 0.2901 & 0.3593 & 0.4261 & 0.5704 & 0.6633 \\
$b335$ & 0.0000 & 0.0004 & 0.0031 & 0.0268 & 0.0853 & 0.1890 & 0.2752 & 0.3535 & 0.4192 & 0.5756 & 0.6634 \\
$b336$ & 0.0000 & 0.0004 & 0.0031 & 0.0272 & 0.0922 & 0.2024 & 0.2902 & 0.3690 & 0.4354 & 0.5826 & 0.6659 \\
$b337$ & 0.0000 & 0.0013 & 0.0050 & 0.0217 & 0.0787 & 0.1824 & 0.2641 & 0.3380 & 0.4097 & 0.5607 & 0.6439 \\
$b338$ & 0.0000 & 0.0013 & 0.0053 & 0.0234 & 0.0750 & 0.1755 & 0.2581 & 0.3317 & 0.4051 & 0.5585 & 0.6427 \\
$b339$ & 0.0000 & 0.0002 & 0.0031 & 0.0170 & 0.0680 & 0.1827 & 0.2808 & 0.3571 & 0.4293 & 0.5779 & 0.6819 \\
$b340$ & 0.0000 & 0.0002 & 0.0031 & 0.0186 & 0.0724 & 0.1840 & 0.2815 & 0.3575 & 0.4295 & 0.5687 & 0.6493 \\
 \end{longtable*}

\end{document}